\documentclass[12pt]{iopart}

\usepackage{hyperref} 
\usepackage{graphicx}
\usepackage{color}
\usepackage{dcolumn}
\usepackage{bm}
\usepackage{verbatim} 
\usepackage{cite} 
\usepackage[normalem]{ulem}

\begin{document}


    \long\def \cblu#1{\color{blue}#1}
    \long\def \cred#1{\color{red}#1}
    \long\def \cgre#1{\color{green}#1}
    \long\def \cpur#1{\color{purple}#1}

\newcommand{\eric}[1]{{\color{blue}#1}}
\newcommand{\guido}[1]{{\color{violet}#1}}
\newcommand{\matthias}[1]{{\color{blue}#1}}
\newcommand{\fabian}[1]{{\color{blue}#1}}
\newcommand{\di}[1]{{\color{blue}#1}}
\newcommand{\ericC}[1]{{\color{red}\textit{\textbf{Eric:} #1}}}
\newcommand{\guidoC}[1]{{\color{red}\textit{\textbf{Guido:} #1}}}
\newcommand{\matthiasC}[1]{{\color{red}\textit{\textbf{Matthias:} #1}}}
\newcommand{\fabianC}[1]{{\color{red}\textit{\textbf{Fabian:} #1}}}
\newcommand{\diC}[1]{{\color{red}\textit{\textbf{Di:} #1}}}

\def\FileRef{
\input FName
{
\newcount\hours
\newcount\minutes
\newcount\min
\hours=\time
\divide\hours by 60
\min=\hours
\multiply\min by 60
\minutes=\time
\
\advance\minutes by -\min
{\small\rm\em\the\month/\the\day/\the\year\ \the\hours:\the\minutes
\hskip0.125in{\tt\FName}
}
}}

\mathchardef\muchg="321D
\let\na=\nabla
\let\pa=\partial

\let\muchg=\gg

\let\t=\tilde
\let\ga=\alpha
\let\gb=\beta
\let\gc=\chi
\let\gd=\delta
\let\gD=\Delta
\let\ge=\epsilon
\let\gf=\varphi
\let\gg=\gamma
\let\gh=\eta
\let\gj=\phi
\let\gF=\Phi
\let\gk=\kappa
\let\gl=\lambda
\let\gL=\Lambda
\let\gm=\mu
\let\gn=\nu
\let\gp=\pi
\let\gq=\theta
\let\gr=\rho
\let\gs=\sigma
\let\gt=\tau
\let\gw=\omega
\let\gW=\Omega
\let\gx=\xi
\let\gy=\psi
\let\gY=\Psi
\let\gz=\zeta

\let\lbq=\label
\let\rfq=\ref
\let\na=\nabla
\def\daI{{\dot{I}}}
\def\dsq{{\dot{q}}}
\def\dgj{{\dot{\phi}}}

\def\bgs{\bar{\sigma}}
\def\bgh{\bar{\eta}}
\def\bgg{\bar{\gamma}}
\def\bgF{\bar{\Phi}}
\def\bgY{\bar{\Psi}}

\def\baF{\bar{F}}
\def\bsj{\bar{j}}
\def\baJ{\bar{J}}
\def\bsp{\bar{p}}
\def\baP{\bar{P}}
\def\bsx{\bar{x}}

\def\hgj{\hat{\phi}}
\def\hgq{\hat{\theta}}

\def\HaT{\hat{T}}
\def\HaR{\hat{R}}
\def\Hsb{\hat{b}}
\def\Hsh{\hat{h}}
\def\Hsz{\hat{z}}

\let\gG=\Gamma
\def\taA{{\tilde{A}}}
\def\taB{{\tilde{B}}}
\def\taG{{\tilde{G}}}
\def\tsp{{\tilde{p}}}
\def\tsv{{\tilde{v}}}
\def\tgF{{\tilde{\Phi}}}

\def\wgx{{\bm{\xi}}}
\def\wgz{{\bm{\zeta}}}

\def\wse{{\bf e}}
\def\wsk{{\bf k}}
\def\wsi{{\bf i}}
\def\wsj{{\bf j}}
\def\wsl{{\bf l}}
\def\wsn{{\bf n}}
\def\wsr{{\bf r}}
\def\wsu{{\bf u}}
\def\wsv{{\bf v}}
\def\wsx{{\bf x}}

\def\vaB{\vec{B}}
\def\vse{\vec{e}}
\def\vsh{\vec{h}}
\def\vsl{\vec{l}}
\def\vsv{\vec{v}}
\def\vgn{\vec{\nu}}
\def\vgk{\vec{\kappa}}
\def\vgt{\vec{\gt}}
\def\vgx{\vec{\xi}}
\def\vgz{\vec{\zeta}}

\def\waA{{\bf A}}
\def\waB{{\bf B}}
\def\waD{{\bf D}}
\def\waE{{\bf E}}
\def\waJ{{\bf J}}
\def\waV{{\bf V}}
\def\waX{{\bf X}}

\def\R#1#2{\frac{#1}{#2}}
\def\btbl{\begin{tabular}}
\def\etbl{\end{tabular}}
\def\bqbl{\begin{eqnarray}}
\def\eqbl{\end{eqnarray}}
\def\ebox#1{
  \begin{eqnarray}
    #1
\end{eqnarray}}


\def \cred#1{{\color{red}(#1)}}
\def \cblu#1{{\color{blue}#1}}

\title[Particle-based JOREK impurity model]{Collisional-radiative non-equilibrium impurity treatment for JOREK simulations}
\author{D. Hu$^{1}$, G.T.A. Huijsmans$^{2,3}$, E. Nardon$^2$, M. Hoelzl$^4$, M. Lehnen$^5$, D. Bonfiglio$^6$ \& JOREK Team\footnote{See the author list of Ref \cite{Hoelzl2021NF} for a list of current team members}}
 \ead{hudi2@buaa.edu.cn}
\address{
$^1$Beihang University, No. 37 Xueyuan Road, Haidian District, 100191 Beijing, China.
}
\address{
$^2$CEA, IRFM, F-13108 Saint-Paul-Lez-Durance, France
}
\address{
$^3$Eindhoven University of Technology, De Rondom 70 5612 AP Eindhoven, the Netherlands.
}
\address{
$^4$Max Planck Institute for Plasma Physics, Boltzmannstr. 2, 85748 Garching b. M., Germany
}
\address{
$^5$ITER Organization, Route de Vinon sur Verdon, CS 90 046,13067 Saint Paul-lez-Durance, Cedex, France.
}
\address{
$^6$Consorzio RFX-CNR, ENEA, INFN, Universit\`{a} di Padova, Acciaierie Venete SpA. I-35127 Padova, Italy.
}
\ead{hudi2@buaa.edu.cn}

\vspace{10pt}
\begin{indented}
\item[]\today
\end{indented}

\begin{abstract}
A collisional-radiative non-equilibrium impurity treatment for JOREK 3D nonlinear magneto-hydrodynamic (MHD) simulations has been developed. The impurities are represented by super-particles flowing along the fluid velocity field lines, while ionizing and recombining independently according to ADAS data and local fluid density and temperature. The non-equilibrium impurity contributions are then projected back to the fluid field for self-consistent time evolution. A 2D test case is used to compare the new non-equilibrium impurity model against previous Coronal Equilibrium (CE) impurity treatment, as well as to compare the non-equilibrium impurity behavior between the single and the two temperature model. Further, we conduct benchmark with previously published coronal non-equilibrium results by other 3D nonlinear MHD codes such as M3D-C1 and NIMROD. The new non-equilibrium treatment is shown to successfully capture the early phase cooling by weakly ionized impurities which the CE model missed. The benchmarks with M3D-C1 and NIMROD show general agreement in both the integrated quantities and the 2D profile evolution, despite the difference in the atomic model used. The above comparison and benchmark cases demonstrate the capability of the non-equilibrium impurity model for JOREK, paving the way for more sophisticated 3D non-linear Massive Material Injection (MMI) simulations which have important applications in disruption mitigation studies.
\end{abstract}

%
%
%
\maketitle
%
%

\section{Introduction}
\label{s:Intro}

Unmitigated disruptions could represent a severe threat to future high performance tokamaks such as ITER \cite{Lehnen2015JNM}, thus effective and efficient Disruption Mitigation Systems (DMSs) are essential for the sustained operation of such devices. The injection of massive amounts of impurity atoms or hydrogen isotopes are the most promising techniques currently in development \cite{Hollmann2015POP,Baylor2015FST}, and there have been extensive experimental explorations concerning Massive Gas Injection (MGI) \cite{Lehnen2011NF,Reux2015JNM,Pautasso2011NF,
Hollmann2011JNM,Duan2017NF,Chen2018NF,Ding2018PST}, Shattered Pellet Injection (SPI) \cite{Shiraki2018NF,Raman2020NF,Sweeney2020NF,
Xu2018FST,Li2018RSI,Baylor2019NF,Park2020FED} as well as the newly emerged Dispersive Shell Pellet Injection (DSPI) techniques \cite{Hollmann2019PRL}. For the better understanding of the aforementioned injection dynamics, numerous MGI simulations have been carried out \cite{Izzo2008POP,Izzo2013POP,Izzo2015NF,Fil2015POP,Nardon2017PPCF,Zafar2021PST}, while explorations into SPI \cite{Di2018NF,Kim2019POP,Hoelzl2020POP,Nardon2020NF,Di2021NF} and DSPI \cite{Izzo2020NF} are also being conducted by 3D nonlinear magneto-hydrodynamic (MHD) codes such as JOREK, M3D-C1 and NIMROD which provide important insight on the interplay between the MHD modes and the injected material. Understanding such interplay is instrumental for obtaining deeper knowledge of the MHD destabilization and the particle transport after the injections, and ultimately important to the endeavor of finding the most efficient DMS configuration.

The JOREK code is a 3D nonlinear code with both reduced MHD and full MHD capability \cite{Hoelzl2021NF}. It has been used to investigate impurity Massive Material Injection (MMI) dynamics \cite{Di2021NF,Nardon2021PPCF} assuming a Coronal Equilibrium (CE) distribution of the charge states \cite{Mosher1974PRA}. Such assumption, however, is not always accurate in a massive injection scenario, especially in the early phase of the injection and the cooling.
Indeed, the timescale of the injection and the transport during this phase could be comparable with the timescale needed for the impurity charge state distribution to reach the equilibrium \cite{Smith2010APJ}, so that the radiation power density predicted by the CE assumption could suffer inaccuracy compared with that of a more detailed non-equilibrium model as is shown in Fig.\,2 of Ref.\,\cite{Di2021NF}. This is especially important if the temperature drop is not fast enough during the early injection period or during the Thermal Quench (TQ) when sudden outgoing heat flux reheats the outer plasma region, since the CE assumption tends to underestimate the line radiation contribution from the weakly ionized states.

To address the aforementioned potential issue, we hereby develop a particle-based collisional-radiative non-equilibrium model for a better treatment of the impurity charge state distribution. The model relies on so-called ``marker-particles'' which flow along the fluid velocity field line while ionizing and recombining independently according to local plasma temperature and their own ionization-recombination probability. The resulting charge state distribution is then projected to the finite element grid and used in the fluid equations being solved. To demonstrate the capability of this new treatment, we will present a set of comparison and benchmark concerning this non-equilibrium impurity treatment.

The rest of the report is arranged as follows. First, we introduce the implementation of the particle-based impurity treatment in Section \ref{s:Implementation} and describe the axisymmetric equilibrium and simulation setup in Section \ref{s:Eq}. We then show the comparison of the particle-based model against the previous CE impurity model in Section \ref{s:ParModvsCEmod}. The comparison between the single temperature model and the two temperature model both utilizing the non-equilibrium impurity treatment are shown in Section \ref{s:TeTiComparison}. Further, we benchmark the particle-based model with previous M3D-C1 and NIMROD axisymmetric simulations of a central massive gas injection in a DIII-D-like plasma with both argon and neon injections in Section \ref{s:Benchmark}. Last, the conclusion is given in Section \ref{s:Conclusion}. We further demonstrate that very good agreement could be reached with the M3D-C1 and NIMROD results when using an older version of the ADAS data in \ref{ap:Atomic}.

\section{The implementation of the particle-based impurity treatment}
\label{s:Implementation}

JOREK has implemented the infrastructure for both test particle and coupled particle-fluid simulations in recent years \cite{VanVugt2018EPS,VanVugt2019POP,VanVugtThesis}. Our non-equilibrium impurity treatment is based on the existing infrastructure, but with some modification to the particle pusher and ionization-recombination process.

In this section, we introduce the implementation of the particle-based impurity model. We first show how the particles are evolved which includes the particle pusher and the evolution of its charge states. We then introduce how the particles contribute to the fluid evolution by their moment projection.

\subsection{Marker-particle pusher and particle generation}
\label{ss:MarkerPartPusher}

The ``marker-particles'' considered in out treatment do not evolve according to their own respective Newton's law, but are simply advected with the fluid velocity field, hence the name. This simplification is in accordance with our previous assumption that the inter-species friction and the charge exchange is strong enough so that all species and their different charge states share the same velocity \cite{Di2021NF}. Currently, the additional diffusive transport of the fluid density is not yet reflected in the particle advection due to its sub-dominant role.

In JOREK, the fluid velocity field can be directly calculated by the flow function and the parallel velocity field:
\bqbl
\lbq{eq:MagField}
\waB
=
F_0\na\gj
+\na\gy\times\na\gj
,\eqbl
\bqbl
\lbq{eq:VelField}
\wsv
=
v_\|\waB
-R^2\na u\times\na \gj
.\eqbl
Here $F_0/R$ is the toroidal magnetic field while $F_0$ is assumed to be constant in space and time. The poloidal magnetic flux is $\gy$, while $v_\|\waB$ is the parallel velocity, $R$ is the major radius, $\gj$ is the toroidal angle and $u$ is the poloidal flow potential.
Using the above expressions, we can write down the velocity as
\bqbl
\lbq{eq:VelFieldExplicit}
\wsv
&
=
&
\left(-R\pa_Zu+\R{v_\|}{R}\pa_Z\gy\right)\wse_{R}
+\left(R\pa_Ru-\R{v_\|}{R}\pa_R\gy\right)\wse_{Z}
+\R{F_0}{R} v_\|\wse_\gj
.\eqbl
The marker-particle pusher advects the marker-particles with the fluid velocity, {\it i.e.} the particle velocity is the same as the local fluid velocity.

In JOREK, the particle time step is set to be a fraction of the fluid time step, so that the fluid fields are evolved one time step after a certain number of particle steps. The pusher hence pushes the marker-particles for several particle time steps under a given fluid field before the fluid field is updated. From past simulations, setting the particle time step to be smaller than $10^{-8}s$ is enough to assure accuracy in the particle evolution \cite{VanVugt2019POP}.

At the end of each fluid time step, a number of marker particles are added to the simulation according to the impurity density source term in the fluid evolution, which in turn is determined by the gas injection source or the pellet ablation. The amplitude and shape of such density source is detailed in Ref.\,\cite{Di2021NF}. The generated super-particles are equal in their weight (the amount of real particles represented by a given super-particle), and their generation location is obtained by the reject-sampling method according to the aforementioned source terms. We assume all particles are generated as neutrals.

\subsection{The ionization, recombination and radiation}
\label{ss:IonRecon}

In a MMI scenario, the total injected atom number could easily be as high as $10^{22}$ impurity atoms for large devices such as ITER. This either demands using a tremendous amount of super-particles, or requires each super-particle to represent a large population of real particles. The former is undesirable in terms of simulation cost, while the latter is detrimental to the smoothness of particle moment projection onto the fluid field. The large super-particle weight could be especially problematic when we are concerned with the dynamic balance of the ionization and the recombination where we have the cancellation of two large terms.

On the other hand, due to the use of the marker-particle pusher, we are free to allocate a charge state density array to each super-particle instead of a single charge number, reflecting the charge state distribution of the real particles represented by that super-particle. This has the benefit of providing a smoother ionization-recombination representation compared with the traditional single charge number treatment, enabling us to simulate a MMI scenario with moderate number of super-particles. Typically, we use $10^7$ super-particles to represent $10^{21}$ to $10^{22}$ impurity atoms, thus the typical weight of marker-particles is $10^{14}$ to $10^{15}$. Empirically, depending on the number of ``active'' super-particles, the additional computational time due to the super-particle evolution ranges from negligible in the early simulations to tens of seconds per fluid time-step late into the simulation. As a reference, the fluid time-step typically takes thirty seconds for an axisymmetric run on four nodes, each with 32 FeiTeng FT-2000+ processors.

The charge state distribution for each super-particle is evolved as the following \cite{Hughes1985ApJ}:
\bqbl
\lbq{eq:Ion-Rec}
\R{d f_i}{dt}
=
n_e\left[
S_{i-1}(T_e)f_{i-1}
-\left(S_{i}(T_e)f_{i}+\ga_{i}(T_e)\right)f_i
+\ga_{i+1}f_{i+1}
\right]
.\eqbl
Here, $S_i$ and $\ga_i$ are the ionization rate and recombination rate of charge state $i$ respectively, while $f_i$ is the number density of that charge state, and $n_e$ is the electron density. Both the ionization and recombination data are obtained from the open ADAS atomic database \cite{OPENADAS}. The open ADAS data is calculated by using the collisional-radiative model, as opposed to the coronal model ADPAK data used in Ref. \cite{Lyons2019PPCF}, which may account for some deviation that we will observe in the following sections. To be specific, we are using the 1989 version of the argon ADAS data and the 1996 version of the neon ADAS data. Indeed, better agreement is found with the M3D-C1 and NIMROD result when older versions of ADAS data are used as shown in \ref{ap:Atomic}. However, due to the availability issue of some of the old atomic data, in the main article we will continue to use the newer versions as specified above. We will also show the comparison of the line radiation, the ionization and the recombination coefficient of neon between the ADAS data and the ADPAK data in \ref{ap:Atomic}.

\subsection{The projection of particle contribution}
\label{ss:PartProj}

The last piece in the JOREK particle-fluid coupling is the projection of particle moment onto the fluid fields. The exact projection method is detailed in Ref.\,\cite{VanVugtThesis}. In our treatment, we mainly project the following five quantities onto the fluid fields.

The first is the ionization power density, which is representing the energy taken from the electron thermal energy to compensate the impurity's ionization energy. At each particle time step as the charge state distribution is evolved, we collect the change in the ionization energy for each super-particle:
\bqbl
\lbq{eq:Eion}
\gD E_{ion}
=
n_e\gD t\sum_{i=0}^{Z-1}\left(
N_{imp}^{i+}S_{i}(T_e)-N_{imp}^{i+1+}\ga_{i}(T_e)
\right)E_{ion}^{i+}
.\eqbl
Here, $N_{imp}^{i+}$ is the number of real particles at charge state $i+$, $E_{ion}^{i+}$ is the ionization energy when ionizing from $i+$ charge state to $i+1+$ charge state, $\gD t$ is the particle time step. This energy change will act on the electron fluid thermal energy as a source-sink term by moment projection. It should be noted that, physically, the ionization energy would be lost by recombination radiation upon recombination. Here in the modelling, we added the ionization energy back to the electron thermal energy to cancel part of the ADAS recombination radiation, to avoid double counting the part of recombination energy that is taken from the ionization potential energy.

The second is the radiation power density. With a given charge state distribution for a given super-particle, the total radiation power $P_{rad}$ is then a function of the electron density and temperature \cite{Summers1979JPB}.
\bqbl
\lbq{eq:Prad}
P_{rad}
&
=
&
n_e n_{imp} L_{rad}(n_e,T_e)
\nonumber
\\
&
=
&
n_e\sum_{i=0}^{Z}n_{imp}^{i+}\left(L_L^{i+}(n_e,T_e)+L_R^{i+}(n_e,T_e)+L_B^{i+}(n_e,T_e)\right)
.\eqbl
Here, $n_{imp}$ is the total density of the impurity species, while $n_{imp}^{i+}$ is that of each charge state. The radiation power function $L_L^{i+}$, $L_R^{i+}$ and $L_B^{i+}$ are the line radiation, the recombination radiation and the bremsstrahlung radiation of each charge state respectively.
The radiation power functions are also obtained from the open ADAS database \cite{OPENADAS}.

The third is the effective charge of the impurity species which is defined as
\bqbl
\lbq{eq:EffectiveCharge}
Z_\textrm{eff}
\equiv
\R{\sum_i{n_iZ_i^2}}{\sum_i{n_iZ_i}}
\nonumber
.\eqbl
Note that the summation here includes all species and their charge states. The effective charge affects the calculation of the Spitzer-like resistivity \cite{Hirshman1978POF}:
\bqbl
\lbq{eq:Resistivity}
\gh
=
Z_\textrm{eff}\R{\gh_0}{\max{\left(T_e,T_{thres}\right)}^{3/2}}\times
\R{1+1.198\,Z_\textrm{eff}+0.222\,Z_\textrm{eff}^2}{1+2.966\,Z_\textrm{eff}+0.753\,Z_\textrm{eff}^2}
.\eqbl
Here we take the cut-off temperature $T_{thres}=1eV$, below which the temperature dependency is not accounted for.

The fourth quantity is the mean charge of the impurity species, which is used to calculate the electron density assuming quasi-neutrality. The last projected quantity is the impurity number density.

\subsection{The governing equations with particle moment contributions}
\label{ss:GovEq}

The formulation of our single and two temperature models under the CE assumption has been detailed in Ref. \cite{Di2021NF}. For the completeness of the paper, we again write down the governing equations with the particle moment contributions in this subsection, but we will only focus on the particle moment contributions instead of going over all the details.

First, we have the induction equation:
\bqbl
\lbq{eq:InductionEq}
\R{\pa\gy}{\pa t}
&
=
&
\gh\left(T_e\right)\gD^*\gy
-R\left\{u,\gy\right\}
-F_0\R{\pa u}{\pa\gj}
,\eqbl
\bqbl
\lbq{eq:AmpereEq}
j
&
=
&
\gD^*\gy
,\quad
j_\gj
=
-j/R
,\eqbl
with Poisson bracket $\left\{f,g\right\}\equiv R\left(\na f\times\na g\right)\cdot\na\gj$. As mentioned in Section \ref{ss:PartProj}, the particle moment contributes to the resistivity through the effective charge $Z_\textrm{eff}$.

Second, the continuity equation for both the total plasma mass density and that for the impurity species is not affected:
\bqbl
\lbq{eq:ContinuityEq}
\R{\pa \gr}{\pa t}
&
=
&
-\na\cdot\left(\gr\wsv\right)
+\na\cdot\left(D\na\gr\right)
+S_{bg}
+S_{imp}
,\eqbl
\bqbl
\lbq{eq:ImpEq}
\R{\pa \gr_{imp}}{\pa t}
=
-\na\cdot\left(\gr_{imp}\wsv\right)
+\na\cdot\left(D\na\gr_{imp}\right)
+S_{imp}
.\eqbl
Here, the impurity density source $S_{imp}$ determines the particle generation by reject-sampling as mentioned in Section \ref{ss:MarkerPartPusher}. We will show later in Section \ref{s:ParModvsCEmod} that the fluid impurity density evolution described here are mostly consistent with that from the particle moment projection before the plasma termination.

Third, we have the perpendicular and the parallel momentum equations:
\bqbl
\lbq{eq:VorticityEq}
R\na\cdot\left[R^2\R{\pa}{\pa t}\left(\gr\na_{pol}u\right)\right]
&
=
&
\R{1}{2}\left\{R^2\left|\na_{pol}u\right|^2,R^2\gr\right\}
+\left\{R^4\gr\gw,u\right\}
\nonumber
\\
&&
-R\na\cdot\left[R^2\na_{pol}u\na\cdot\left(\gr\wsv\right)\right]
+\left\{\gy,j\right\}
\nonumber
\\
&
&
-\R{F_0}{R}\R{\pa j}{\pa\gj}
+\left\{P,R^2\right\}
+R\gm_\bot\left(T_e\right)\na_{pol}^2\gw
,\eqbl
\bqbl
\lbq{eq:VorticityDef}
\gw
&
=
&
\R{1}{R}\R{\pa}{\pa R}\left(R\R{\pa u}{\pa R}\right)
+\R{\pa^2 u}{\pa Z^2}
,\eqbl
\bqbl
\lbq{eq:ParallelEq}
B^2\R{\pa}{\pa t}\left(\gr v_\|\right)
&
=
&
-\R{1}{2}\gr\R{F_0}{R^2}\R{\pa}{\pa \gj}\left(v_\|B\right)^2
-\R{\gr}{2R}\left\{B^2v_\|^2,\gy\right\}
-\R{F_0}{R^2}\R{\pa P}{\pa \gj}
\nonumber
\\
&
&
+\R{1}{R}\left\{\gy,P\right\}
-B^2\na\cdot\left(\gr\wsv\right)v_\|
+B^2\gm_\|\na_{pol}^2v_\|
.\eqbl
The vorticity equation Eq.\,(\rfq{eq:VorticityEq}) is obtained by applying $\na\gj\cdot\na\times\left(R^2\cdots\right)$ on both sides of the momentum equation. The total pressure consists of the electron and the ion contribution $P=P_e+P_i$, with $P_i=\left(n_{bg}+n_{imp}\right)T_i$ and $P_e=n_e T_e=\left(n_{bg}+Z_{imp}n_{imp}\right)T_e$. Here, $n_{bg}$ is the number density of the background hydrogen isotopes, $n_{imp}$ is the impurity number density, $Z_{imp}$ is the projected mean charge number. The electron and the ion temperature are represented by $T_e$ and $T_i$ respectively. For the single temperature model, we have $T_e=T_i$.

Last, we write down the pressure equation for the single and the two temperature model separately. For the single temperature model,
\bqbl
\lbq{eq:PressureEq}
\R{\pa P}{\pa t}
&
=
&
-\wsv\cdot\na P
-\gg P\na\cdot\wsv
+\R{\gg-1}{R^2}\gh\left(T_e\right)j^2
+\na\cdot\left(\gk_\bot\na_\bot T
+\gk_\|\na_\| T\right)
\nonumber
\\
&
&
+\left(\gg-1\right)\gm_\|\left[\na_{pol}\left(v_\|B\right)\right]^2
-\left(\gg-1\right)\left(P_{rad}+P_{ion}\right)
\nonumber
\\
&&
+\R{\gg-1}{2}\wsv\cdot\wsv\left(S_{bg}+S_{imp}\right)
.\eqbl
Here, $P_{rad}$ is the projected radiative power. The ionization power is defined by $P_{ion}\equiv \gD E_{ion}/\gD t$, where $\gD E_{ion}$ is the projected ionization energy loss during the fluid time step $\gD t$. For the two temperature model, the equations are:
\bqbl
\lbq{eq:i_TemperatureEq502}
\R{\pa}{\pa t}P_i
&
=
&
-\wsv\cdot \na P_i
-\gg P_i\na\cdot\wsv
+\na\cdot\left(\gk_\bot\na_\bot T_i
+\gk_{i,\|}\na_\| T_i\right)
\nonumber
\\
&&
+\R{\gg-1}{2}\wsv\cdot\wsv\left(S_{bg}+S_{imp}\right)
+\left(\gg-1\right)\gm_\|\left[\na_{pol}\left(v_\|B\right)\right]^2
\nonumber
\\
&&
+\left(n_{bg}+n_{imp}\right)\left(\pa_t T_i\right)_{c,e}
,\eqbl
\bqbl
\lbq{eq:e_TemperatureEq502}
\R{\pa}{\pa t}P_e
&
=
&
-\wsv\cdot \na P_e
-\gg P_e\na\cdot\wsv
+\na\cdot\left(\gk_\bot\na_\bot T_e
+\gk_{e,\|}\na_\| T_e\right)
\nonumber
\\
&&
+\R{\gg-1}{R^2}\gh\left(T_e\right)j^2
-\left(\gg-1\right)\left(P_{rad}+P_{ion}\right)
+n_e\left(\pa_t T_e\right)_{c,i}
.\eqbl
The additional thermalization terms $\left(n_{bg}+n_{imp}\right)\left(\pa_t T_i\right)_{c,e}$ and $n_e\left(\pa_t T_e\right)_{c,i}$ are also dependent on the projected effective charge since the collisional ion-electron thermalization rate is proportional to the charge number squared of the ion species.

Together, Eq.\,\rfq{eq:InductionEq} to Eq.\,\rfq{eq:e_TemperatureEq502} form our governing equations for the non-equilibrium impurity treatment.

\section{The axisymmetric equilibrium and the simulation setup}
\label{s:Eq}

We provide several comparison and benchmark cases using a DIII-D equilibrium which has been used by the published M3D-C1 \& NIMROD benchmark study \cite{Lyons2019PPCF}. The equilibrium corresponds to the DIII-D shot 137611 at 1950ms. The initial plasma consists of pure deuterium with spatially uniform ion and electron density, $n_e=n_i=10^{20}m^{-3}$. In Ref. \cite{Lyons2019PPCF}, both NIMROD and M3D-C1 were running with the single temperature model. As in Ref. \cite{Lyons2019PPCF}, all runs are axisymmetric and nonlinear. Constant thermal conductivity and particle diffusivity are used here for direct comparison with Ref \cite{Lyons2019PPCF}. To be explicit, we use the isotropic density diffusivity $D=10m^2/s$, and the perpendicular thermal diffusivity is set to $100m^2/s$. The parallel thermal diffusivity is set to $10^8m^2/s$. For the two temperature case, the same heat conduction is used for both electrons and ions. The viscosity corresponds to a momentum diffusivity of $100m^2/s$ initially, but with the same temperature dependence of the resistivity so that the magnetic Prandtl number is constant, as opposed to the cases in Ref. \cite{Lyons2019PPCF} where the momentum diffusivity is constant. This deviation does not significantly impact our result, however, due to the lack of strong MHD activities in an axisymmetric modelling.

We use two different kinds of resistivity models as in Ref. \cite{Lyons2019PPCF}, namely the constant resistivity and the Spitzer-like resistivity. For the constant resistivity, the resistivity is fixed to $10^{-5}\gW m$, while for the Spitzer-like resistivity the expression is given in Eq.\,(\rfq{eq:Resistivity}), with $\gh_0$ set to $1.83339\times10^{-8}\gW m$. Note that, as in Ref. \cite{Lyons2019PPCF}, this resistivity corresponds to an enhanced resistivity $2.444$ times higher than the physical one. In our simulations, the temperature dependency of the resistivity is cut off below $T_{thres} = 1eV$ to avoid numerical instability.

\begin{figure*}
\centering
\noindent
\btbl{c}
\parbox{6.5in}{
    \includegraphics[scale=0.45]{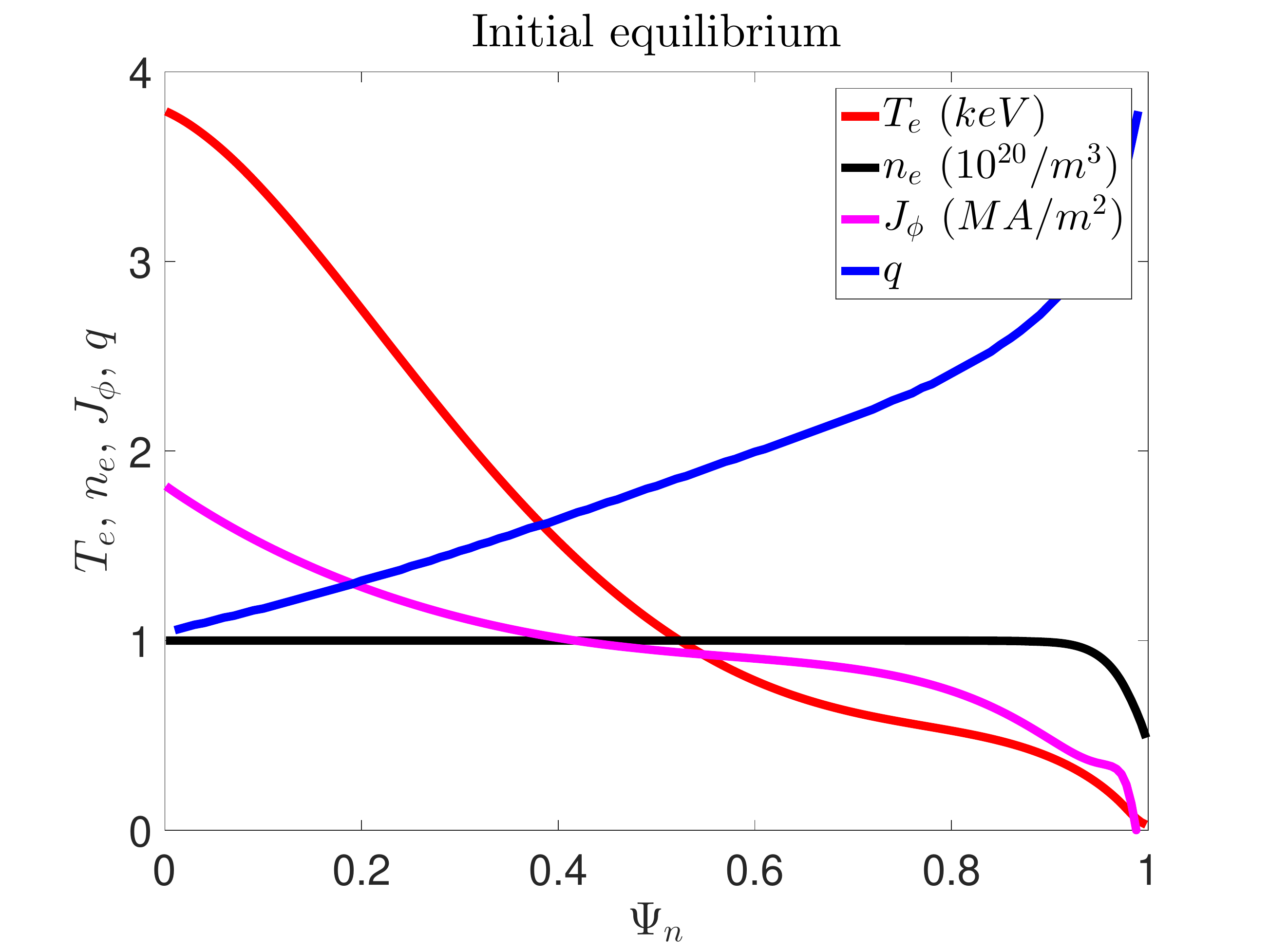}
}
\etbl
\caption{The initial equilibrium. $\gY_N$ is the normalized poloidal flux.}
\label{fig:01}
\end{figure*}

We use a Gaussian shape distribution for our impurity density source with the following shape:
\bqbl
S_n
\propto
\exp{\left(-\R{\left(R-R_{f}\right)^2+\left(Z-Z_{f}\right)^2}{\gD r^{2}_{NG}}\right)}
\times\exp{\left(-\left(\R{\gj-\gj_{f}}{\gD\gj_{NG}}\right)^2\right)}
.\eqbl
In this axisymmetric simulation, we simply set $\gD\gj_{NG}$ to be a very large number, and we set $\gD r^{2}_{NG}=0.356m$, $R_0=1.77037m$ and $Z_0=0.01447m$ so that we have generally the same deposition shape with Ref. \cite{Lyons2019PPCF}. It should be noted that this shape still differs from that is used in Ref. \cite{Lyons2019PPCF} by a $R_0/R$ factor, which would not have any qualitative impact for the plasma dynamics given the geometry concerned. The initial equilibrium profile can be seen in Fig.\,\ref{fig:01}, where the initial electron temperature, electron density, toroidal current density and safety factor profiles are shown as functions of the normalized flux. As in Ref. \cite{Lyons2019PPCF}, the volume integrated impurity atom injection rate is approximately $4.37\times10^{23}$ per second.

\section{Comparison of the particle-based model against the CE fluid model}
\label{s:ParModvsCEmod}

Before we venture to benchmark with the M3D-C1 and the NIMROD results, we first present the comparison between the new particle-based model and the CE model used in Ref. \cite{Di2021NF}. We will be looking at argon injections described by both the single temperature model and the two temperature model, with the Spitzer-like resistivity in both cases. We mainly compare the evolution of the integrated quantities in this section, for a typical 2D profile evolution of the non-equilibrium impurity case please see Fig.\ref{fig:09} and Fig.\ref{fig:10} later on in Section \ref{s:Benchmark}.

\begin{figure*}
\centering
\noindent
\btbl{c}
\parbox{6.5in}{
    \includegraphics[scale=0.45]{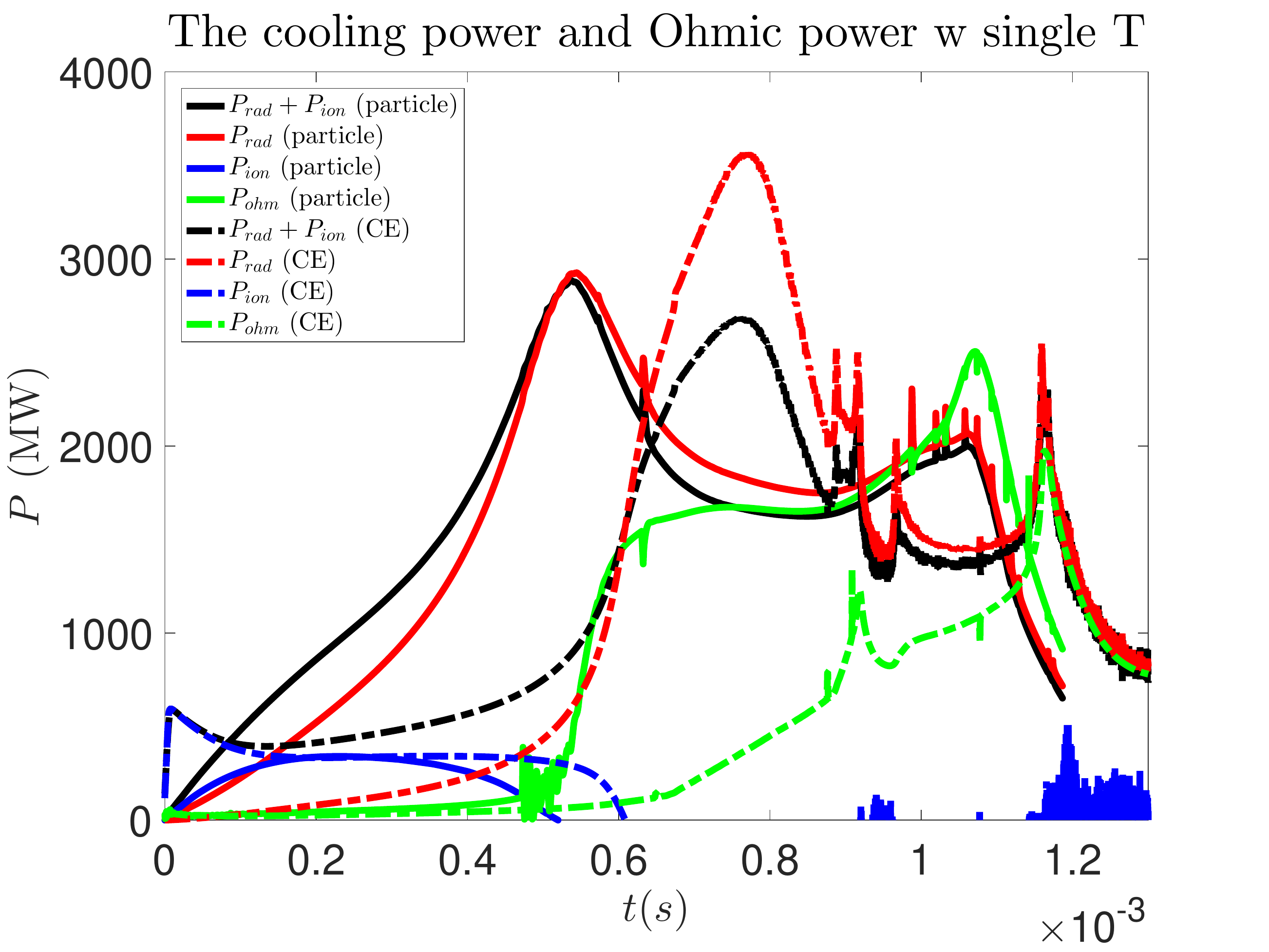}
}
\etbl
\caption{The radiation power (red), ionization power (blue) and Ohmic power (green) for the particle model (solid lines) and the CE model (dash-dot lines) with single temperature treatment. The black solid and dash-dot lines are the combined radiation and ionization power for the particle model and the CE model respectively.}
\label{fig:02}
\end{figure*}

For the single temperature model, the comparison between the particle-based treatment and the CE treatment is shown in Fig.\,\ref{fig:02}. The radiation power from both cases are shown in red, the ionization power in blue and the Ohmic heating power in green. The black lines represent the combined radiation and ionization power. The particle-based results are shown in solid lines while the CE ones are shown in dash-dot lines. The noisiness in the CE ionization power curve is due to the numerical noise when we reconstruct the ionization power from the total ionization energy by finite temporal difference. It can be seen that although the CE model initially start with larger ionization cooling due to the impurities jumping to the equilibrium charge distribution, overall the combined cooling power of the CE model exhibit slower increase compared with the non-equilibrium one. This causes the CE model to underestimate the initial cooling, and the combined cooling power evolution exhibits some delay relative to the non-equilibrium case as can be seen by comparing the black solid and dash-dot lines in Fig.\,\ref{fig:02}, although the peak cooling power is comparable between the two models. A notable deviation is that the non-equilibrium case quickly reached the balance between the combined cooling power and the Ohmic heating at $t\approx 0.7ms$, while the CE case only does so very late into the simulation.

\begin{figure*}
\centering
\noindent
\btbl{c}
\parbox{6.5in}{
    \includegraphics[scale=0.45]{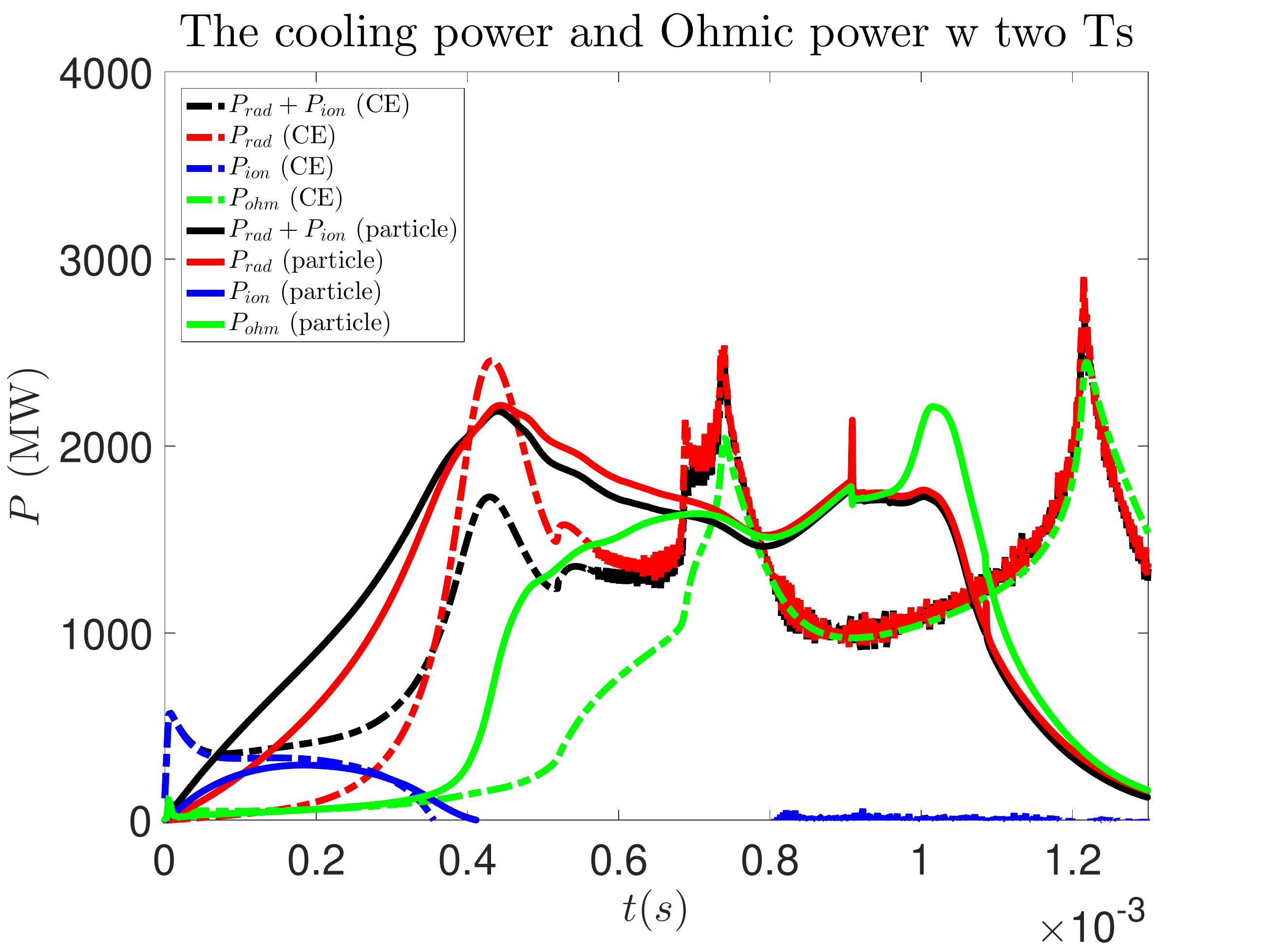}
}
\etbl
\caption{The radiation power (red), ionization power (blue) and Ohmic power (green) for the particle model (solid lines) and the CE model (dash-dot lines) with two temperature treatment. The black solid and dash-dot lines are the combined radiation and ionization power for the particle model and the CE model respectively.}
\label{fig:03}
\end{figure*}

For the two temperature model, a similar deviation between the particle-based model and the CE model is observed as shown in Fig.\,\ref{fig:03}, with the same color scheme as Fig.\,\ref{fig:02}. Again, the solid lines correspond to that of the particle-based model, and the dash-dot lines correspond to that of the CE one. Similar to the single temperature case, the CE model exhibits a stronger ionization cooling due to the immediate jump of the impurity ions to the equilibrium charge state distribution, but the overall cooling power in the early injection phase is underestimated compared with that of the non-equilibrium model, only reaching similar intensity when the plasma is cooled down significantly. A similar delay in the rise of the Ohmic power between the CE model and the non-equilibrium model is also observed. Different from the single temperature model shown in Fig.\,\ref{fig:02}, it can be seen that after a while the radiation and ionization power are balanced by the Ohmic power for both cases, although this happens slightly later for the CE model. The better CE model behavior in the two temperature case probably is due to the fact that the electron temperature is cooled down faster compared with that of the single temperature case. Both the peak cooling power and the peak Ohmic power are comparable for the CE and non-equilibrium models.

\begin{figure*}
\centering
\noindent
\btbl{ccc}
\parbox{2.0in}{
    \includegraphics[scale=0.2]{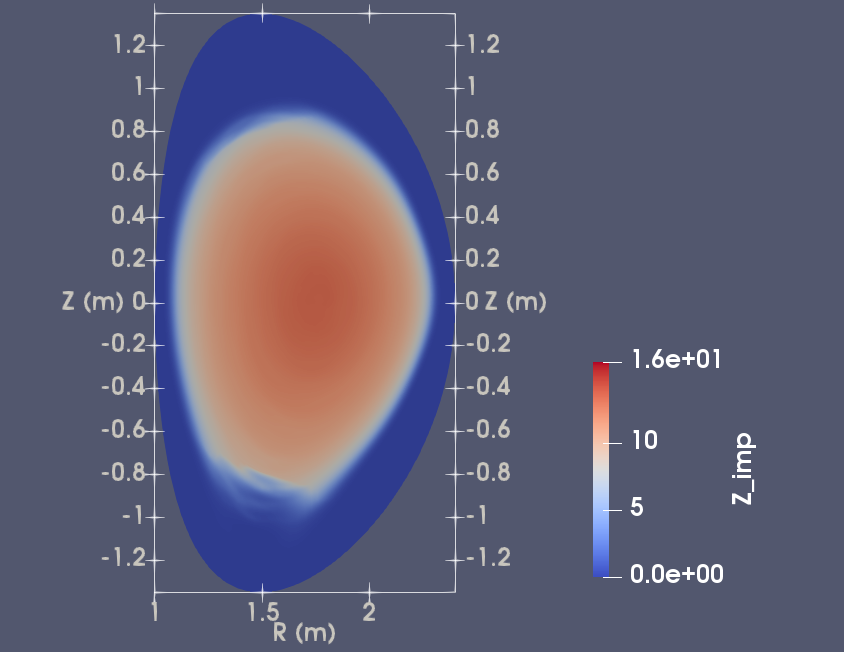}
}
&
\parbox{2.0in}{
	\includegraphics[scale=0.2]{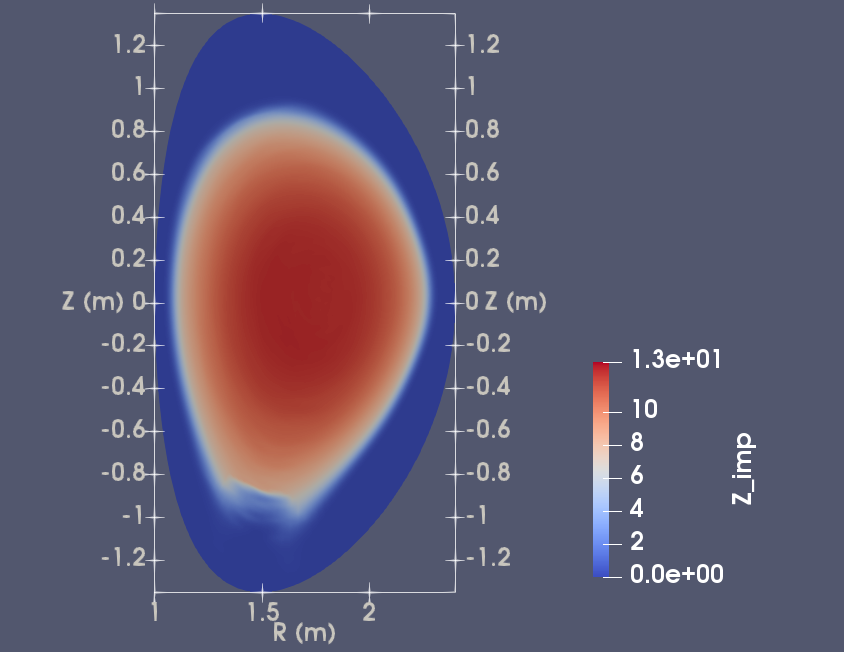}
}
&
\parbox{2.0in}{
	\includegraphics[scale=0.2]{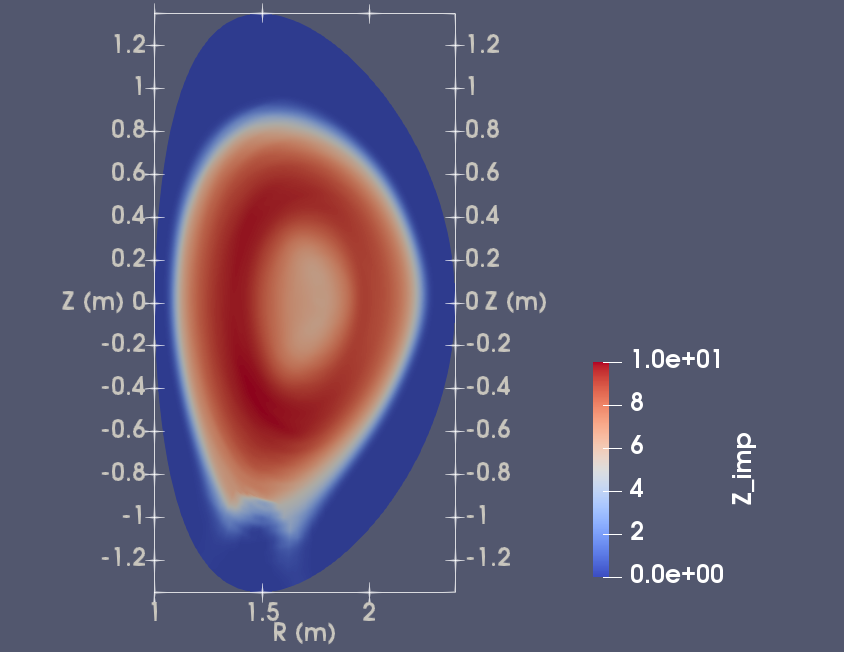}
}
\\
(a)&(b)&(c)
\\
\parbox{2.0in}{
    \includegraphics[scale=0.2]{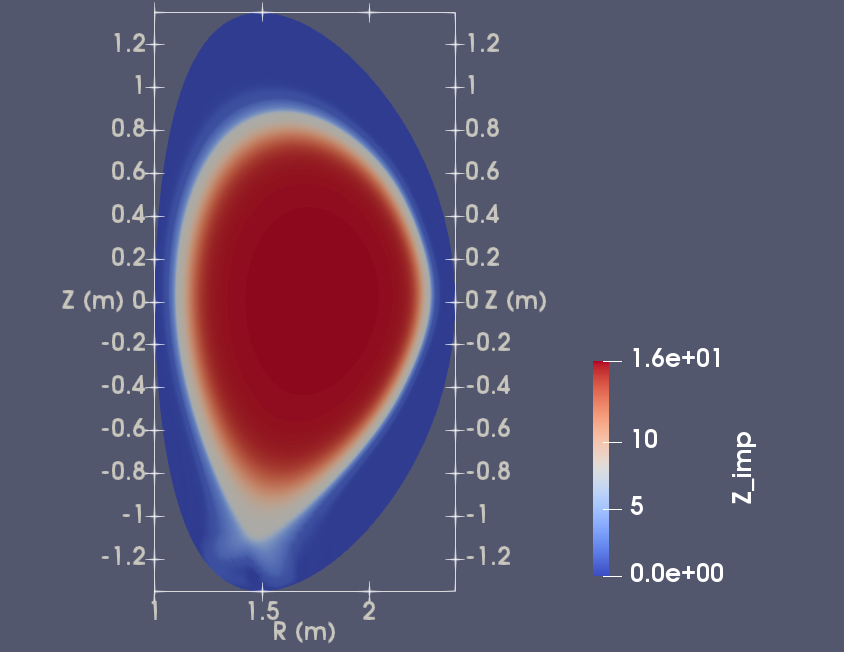}
}
&
\parbox{2.0in}{
	\includegraphics[scale=0.2]{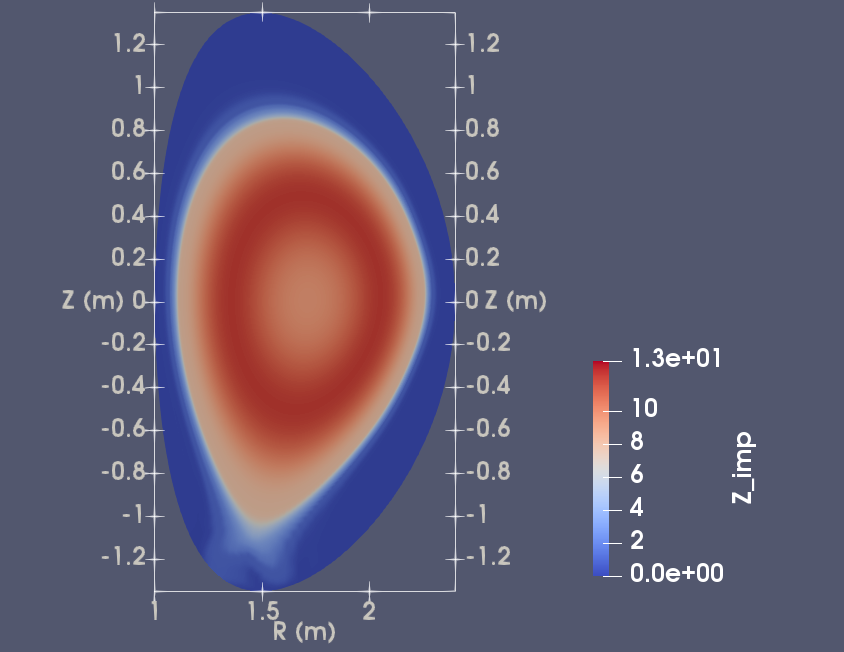}
}
&
\parbox{2.0in}{
	\includegraphics[scale=0.2]{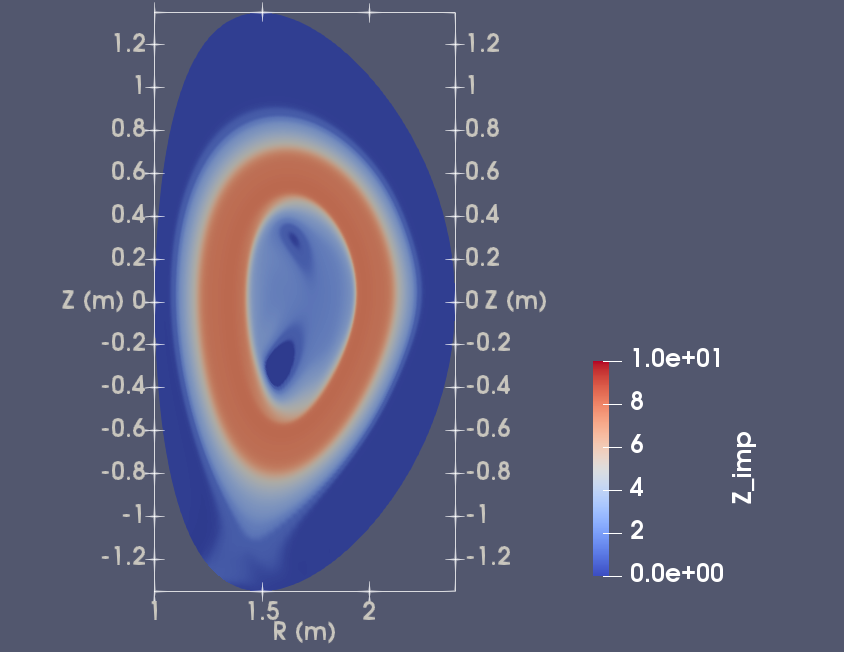}
}
\\
(d)&(e)&(f)
\etbl
\caption{The mean impurity charge for the non-equilibrium treatment at time (a) $t=0.26ms$, (b) $t=0.45ms$ and (c) $t=0.78ms$. For comparison, the corresponding CE mean impurity charge at the same time with (a)-(c) are shown in (d)-(f) respectively. Note the different color scale used in some figures.}
\label{fig:Z_imp_compare}
\end{figure*}

To demonstrate the deviation of the impurity charge state distribution from the CE, we now show the mean impurity charge profile at three different times during the non-equilibrium two temperature argon injection with Spitzer-like resistivity in Fig.\,\ref{fig:Z_imp_compare}(a)-(c). For comparison, we also present the CE mean impurity charge profile for the same run at the same times in Fig.\,\ref{fig:Z_imp_compare}(d)-(f). It is important to note that, Fig.\,\ref{fig:Z_imp_compare}(d)-(f) show the CE mean impurity charge value calculated from the temperature and the density profile of the non-equilibrium impurity simulation, instead of that from a CE simulation. In this way we can directly show how much deviation we have between the non-equilibrium charge number to its equilibrium value. At time $t=0.26ms$, Fig.\,\ref{fig:Z_imp_compare}(a) and (d) correspond to the early phase of the injection, and it is clear that the mean impurity charge is low compared with its equilibrium value due to the existence of large weakly ionized population. At time $t=0.45ms$ which coincides with the first ``turning point'' in Fig.\,\ref{fig:03}, the non-equilibrium result mostly agrees with the CE result, although the CE result already begins to drop in the plasma core due to the central cooling as is shown in Fig.\,\ref{fig:Z_imp_compare}(e) . This is not yet reflected in Fig.\,\ref{fig:Z_imp_compare}(b) due to the fastness of the thermal collapse. Finally, at time $t=0.78ms$, we see that the inside-out cooling results in a halo of strongly ionized impurities under the CE assumption, as is shown in Fig.\,\ref{fig:Z_imp_compare}(f). Such hollow behavior also exists in the non-equilibrium result shown in Fig.\,\ref{fig:Z_imp_compare}(c), but to a lesser extent. The non-equilibrium result also shows higher ionization level since it takes some time to recombine and reach the dropping CE ionization level.

\begin{figure*}
\centering
\noindent
\btbl{ccc}
\parbox{2.0in}{
    \includegraphics[scale=0.2]{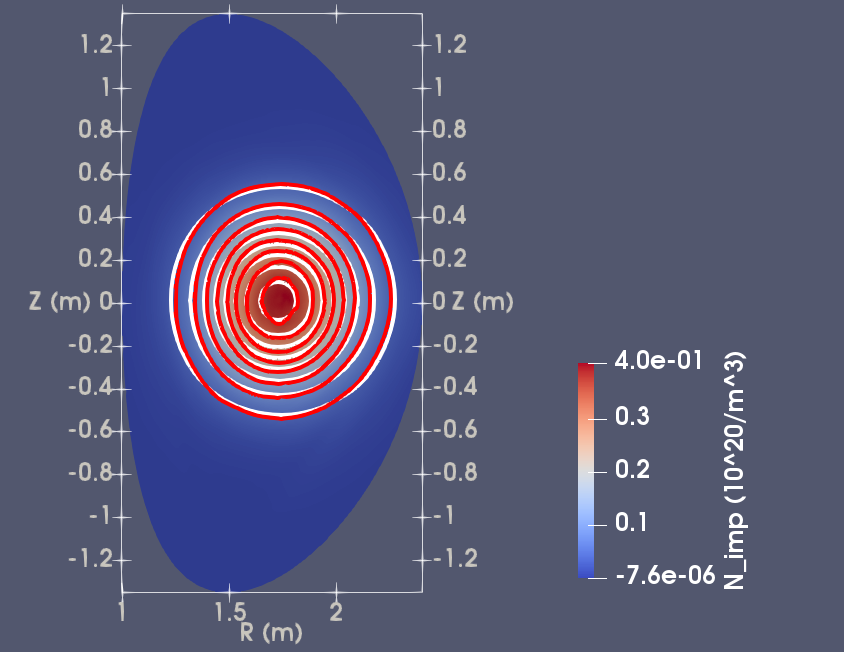}
}
&
\parbox{2.0in}{
	\includegraphics[scale=0.2]{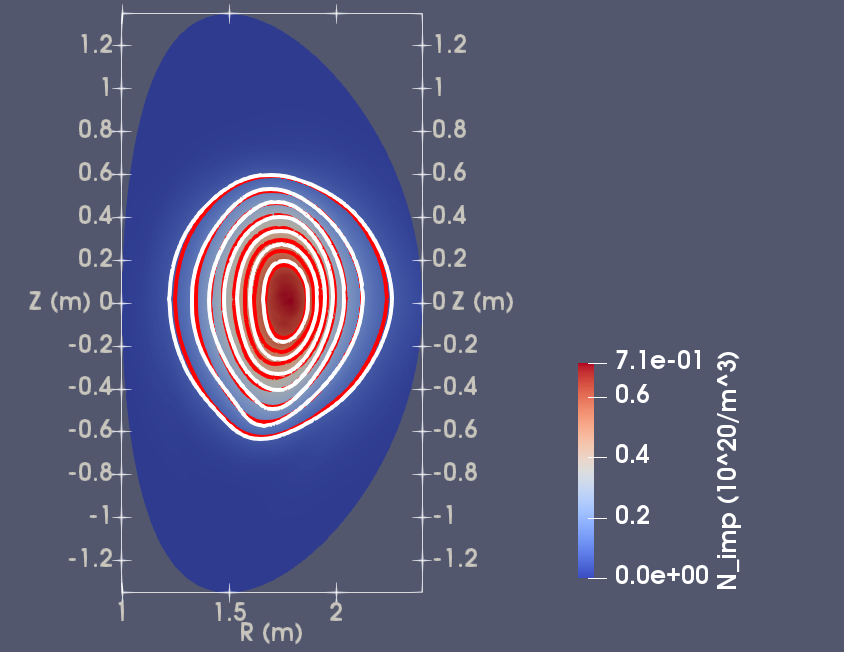}
}
&
\parbox{2.0in}{
	\includegraphics[scale=0.2]{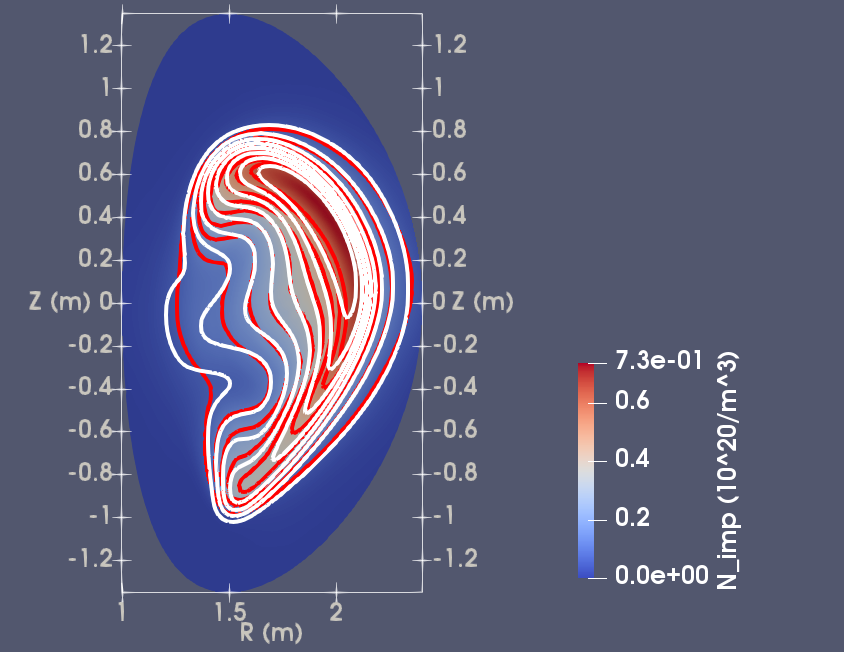}
}
\\
(a)&(b)&(c)
\etbl
\caption{The impurity density contour comparison between the fluid description (white contours) and the particle moment projection (red contours) at (a) $t=0.45ms$, (b) $t=0.78ms$ and (c) $t=1.10ms$. Note that the contours from each color are on exactly the same scale, so that matching contour lines indicate quantitative match of the two representations.}
\label{fig:N_imp_compare}
\end{figure*}

Since we now have two representations of the impurity density profile, one from the fluid description and the other from the projection of particle moments, it is of interest to compare the consistency between the two. The contour comparison of the two temperature argon injection with Spitzer-like resistivity is shown in Fig.\,\ref{fig:N_imp_compare}. In the figure, the red contours corresponds to that from the particle moment projection, while the white ones are from the fluid representation. The contours from each color are on exactly the same level, so that the matching of the contours between the two colors indicate quantitative agreement between the two representations. It is seen that, throughout the majority of the time evolution, good agreement could be found between the two representations as is shown in Fig.\,\ref{fig:N_imp_compare}(a) and (b). Later on at the time of the plasma termination at $t=1.10ms$, some local deviation could be seen, especially in the region with elongated density profile, likely due to the lack of the diffusion contribution in our current particle pusher. We are working on a better pusher to include this effect in the near future. Meanwhile, we do not expect such effect to cause significant impact in future 3D simulations since the stochastic convection along the field line would dominate over the normal diffusion process.

\begin{figure*}
\centering
\noindent
\btbl{c}
\parbox{6.5in}{
    \includegraphics[scale=0.45]{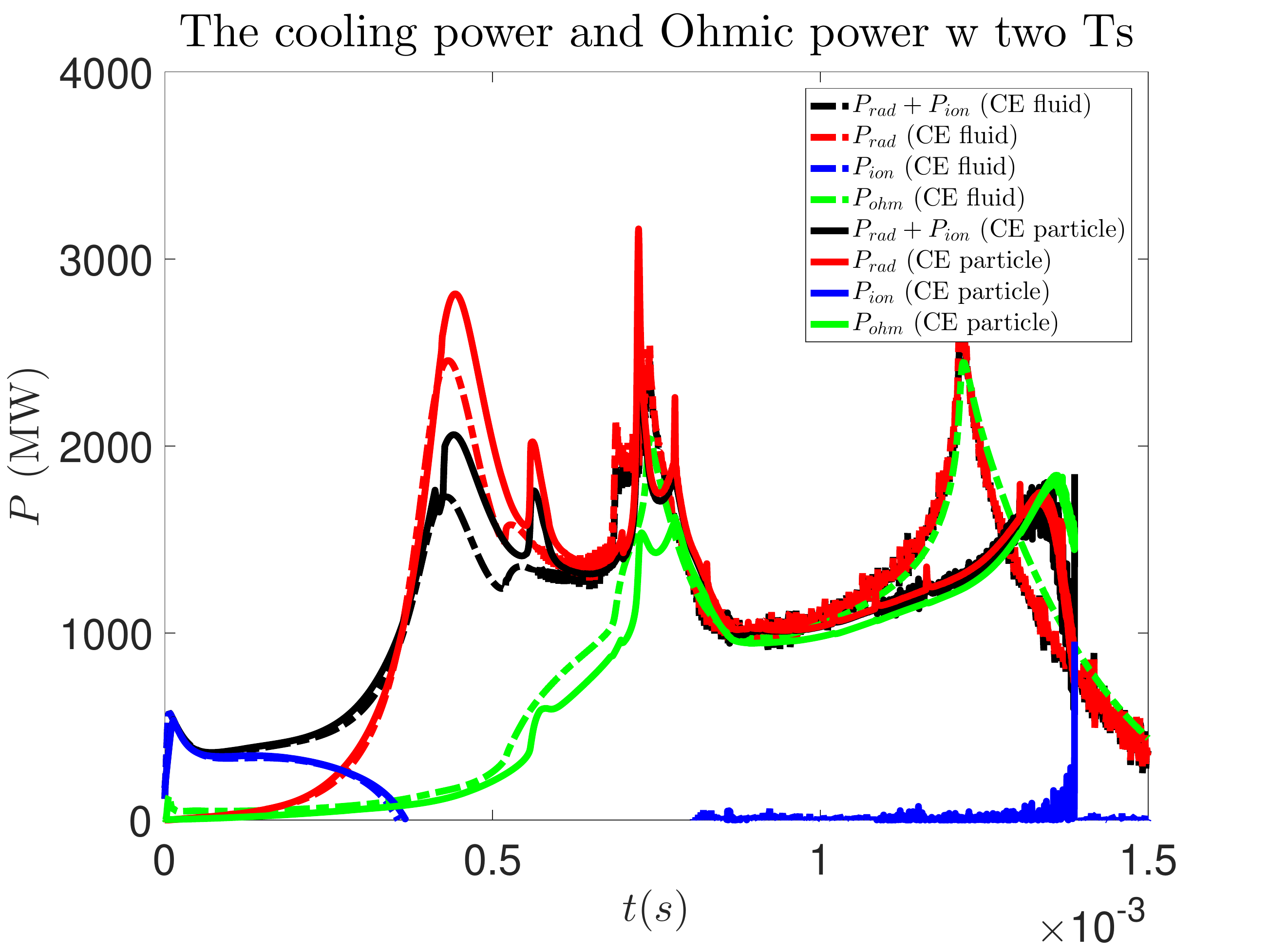}
}
\etbl
\caption{The radiation power (red), ionization power (blue) and Ohmic power (green) for the CE particle model (solid lines) and the CE fluid model (dash-dot lines) with two temperature treatment for argon injection. The black solid and dash-dot lines are the combined radiation and ionization power for the particle model and the CE model respectively.}
\label{fig:CE_particle}
\end{figure*}

Another interesting comparison could be made between the particle treatment and the CE fluid treatment if we enforce the CE charge state distribution upon the super-particles, instead of letting them evolve independently. This could serve to demonstrate the difference between particle representation and the fluid representation directly. A comparison of the cooling power and the ohmic power for the so-called ``CE particle'' against the normal CE fluid representation can be seen in Fig.\,\ref{fig:CE_particle}. Apart from some delay in time close to the plasma termination, it can be seen that very good agreement is found between the CE particle (solid lines) and the CE fluid (dash-dot lines) representation, both in terms of the time of signature events and in terms of the peak value of the cooling power and the ohmic heating. The late stage deviation could be due to the slight deviation in the particle and fluid impurity density representation as previously discussed in Fig.\ref{fig:N_imp_compare}.

All in all, the non-equilibrium model shows a stronger early phase cooling than the CE model. This discrepancy is due to the deviation of impurity charge state distribution from the CE early in the injection. The peak cooling power and Ohmic heating power are comparable between the two models, however the CE model exhibits a delay in reaching those peaks due to the aforementioned inaccuracy in the early cooling phase. In both cases, the total cooling power becomes balanced with the Ohmic power after some time, although the non-equilibrium model reaches this balance faster due to previously explained reasons. The non-equilibrium treatment also shows delayed charge state evolution compared with CE as would be expected. Good agreement is found between the fluid and the particle representation of the impurity density for the majority of the plasma evolution.

\section{Comparison between the single and the two temperature cases}
\label{s:TeTiComparison}

In this section, we use the axisymmetric DIII-D non-equilibrium impurity simulation to demonstrate the difference in behaviours between the single and the two temperature cases.

\begin{figure*}
\centering
\noindent
\btbl{c}
\parbox{6.5in}{
    \includegraphics[scale=0.45]{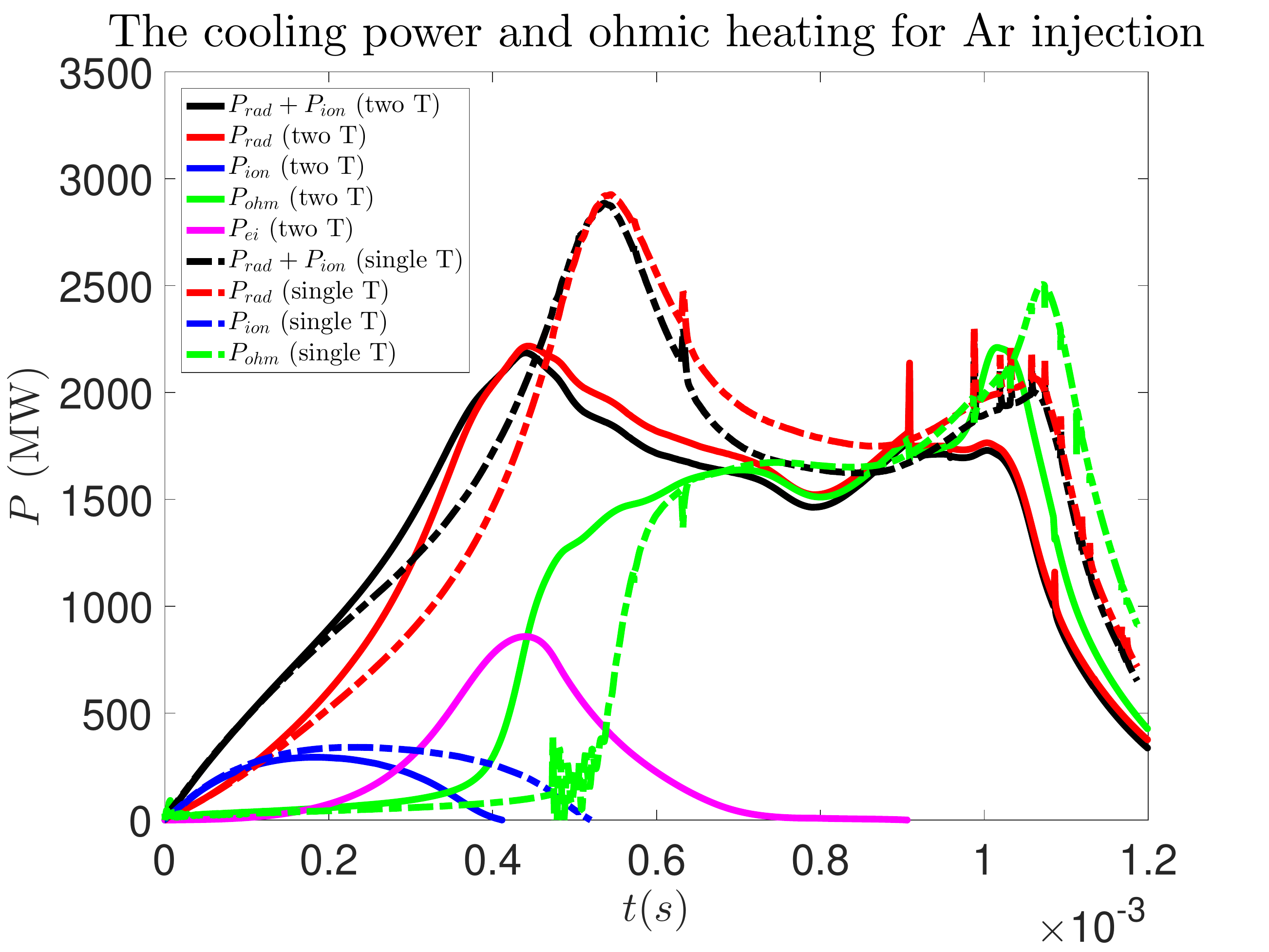}
}
\etbl
\caption{The comparison of JOREK non-equilibrium treatment under the single temperature (dash-dot lines) and the two temperature (solid lines) case. The red lines are the radiation power, the blue ones the ionization power, the black ones represent the above two combined. The Ohmic power is shown in green and the ion-electron energy transfer for the two temperature model is shown in magenta. Apart from the magenta line, this figure is essentially the overlay of the non-equilibrium curves from Fig.\ref{fig:02} and Fig.\ref{fig:03}.}
\label{fig:12}
\end{figure*}

We compare the time evolution of integrated quantities during an axisymmetric DIII-D argon injection with non-equilibrium impurity treatment in Fig.\,\ref{fig:12}. The solid lines indicates the quantities of the two temperature model while the dash-dot lines indicate that of the single temperature model. The red lines are the radiation power, the blue lines are the ionization power and the green lines are the Ohmic power. For the two temperature model, the magenta solid line indicates the thermal energy transfer rate between the ions and the electrons, where a positive value indicates that the ions are transferring energy to the electrons. The black lines indicate the combination of the radiation power and the ionization power.

\begin{figure*}
\centering
\noindent
\btbl{c}
\parbox{6.5in}{
    \includegraphics[scale=0.45]{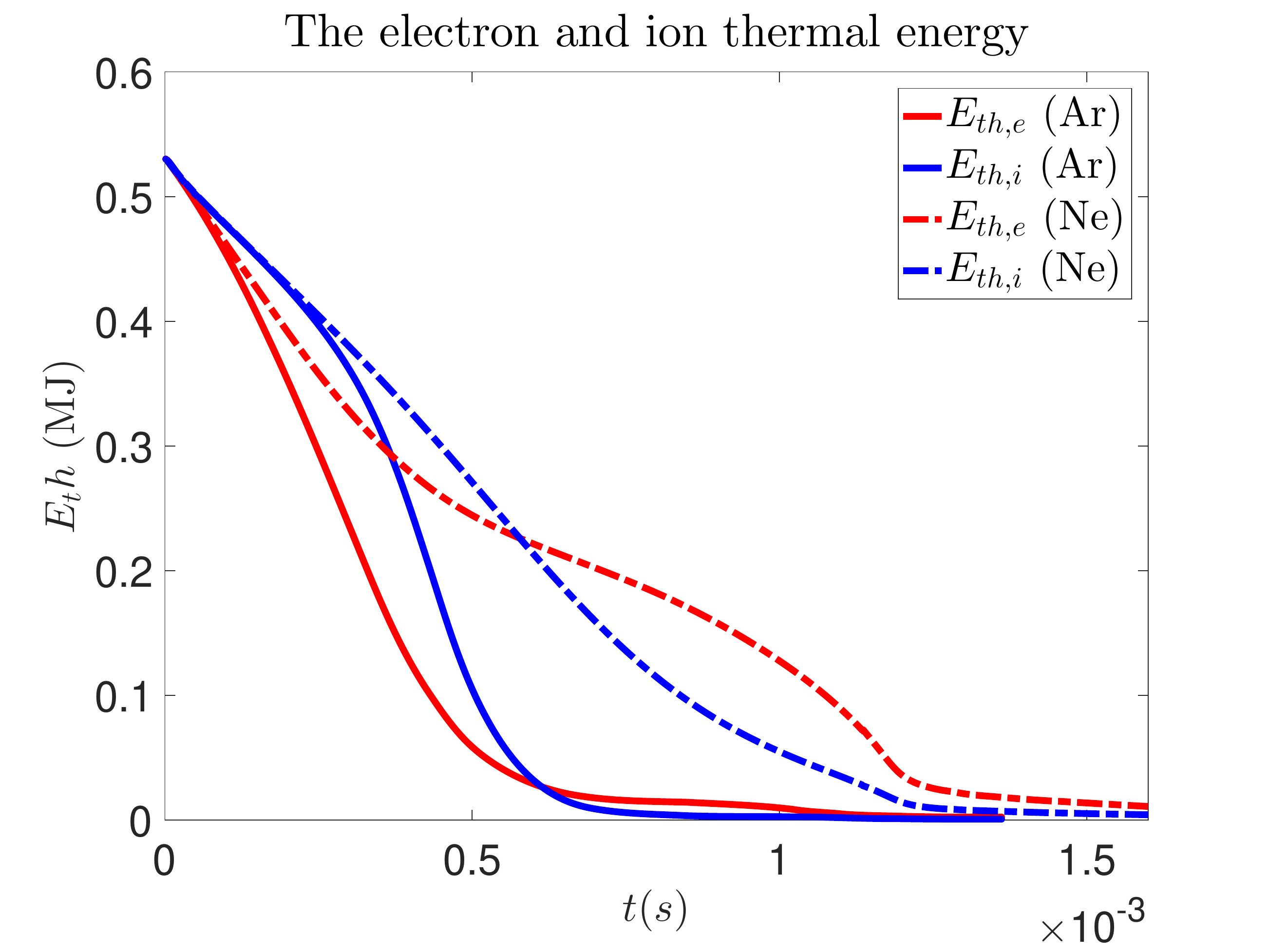}
}
\etbl
\caption{The time evolution of the ion and the electron thermal energy in simulations with the two-temperature model.}
\label{fig:14}
\end{figure*}

The first peak in the cooling power indicates the plasma is cooled down below $100eV$, thus this ``turning point'' coincides with the rise of the Ohmic power. This is also when the ion-electron temperature difference is largest, as suggested by the peak in the ion-electron energy transfer. Then the cooling power and the Ohmic power begin to balance each other, resulting in the plateau region. At this time the inter-species temperature difference is greatly diminished due to thermalization, as indicated by the magenta solid line going to zero. We also see later in Fig.\,\ref{fig:14} that by this time, most of the thermal energy is already depleted and it is the Ohmic conversion of the magnetic energy that is balancing the radiative loss, in agreement with Fig.\,\ref{fig:12} here. Finally there is another peak in the Ohmic power and the cooling power before the plasma terminates.
It can be seen that the two models generally exhibit the same trend of evolution. Both models show the rise of the combined cooling power before the plasma is significantly cooled. Both models show the rise of the Ohmic heating coincides with the turning point of the combined cooling power. Both models exhibit the balance between the Ohmic heating and the radiative cooling, and both models show the spike in the Ohmic heating and the radiative cooling just before the plasma terminates towards the end of the simulations. The two temperature model shows a somewhat reduced radiation peak, as well as an earlier rise of the Ohmic power. This is due to the deviation between the ion and electron temperature in the two temperature case, as opposed to the single temperature case where the two species thermalize immediately.

\begin{figure*}
\centering
\noindent
\btbl{c}
\parbox{6.5in}{
    \includegraphics[scale=0.45]{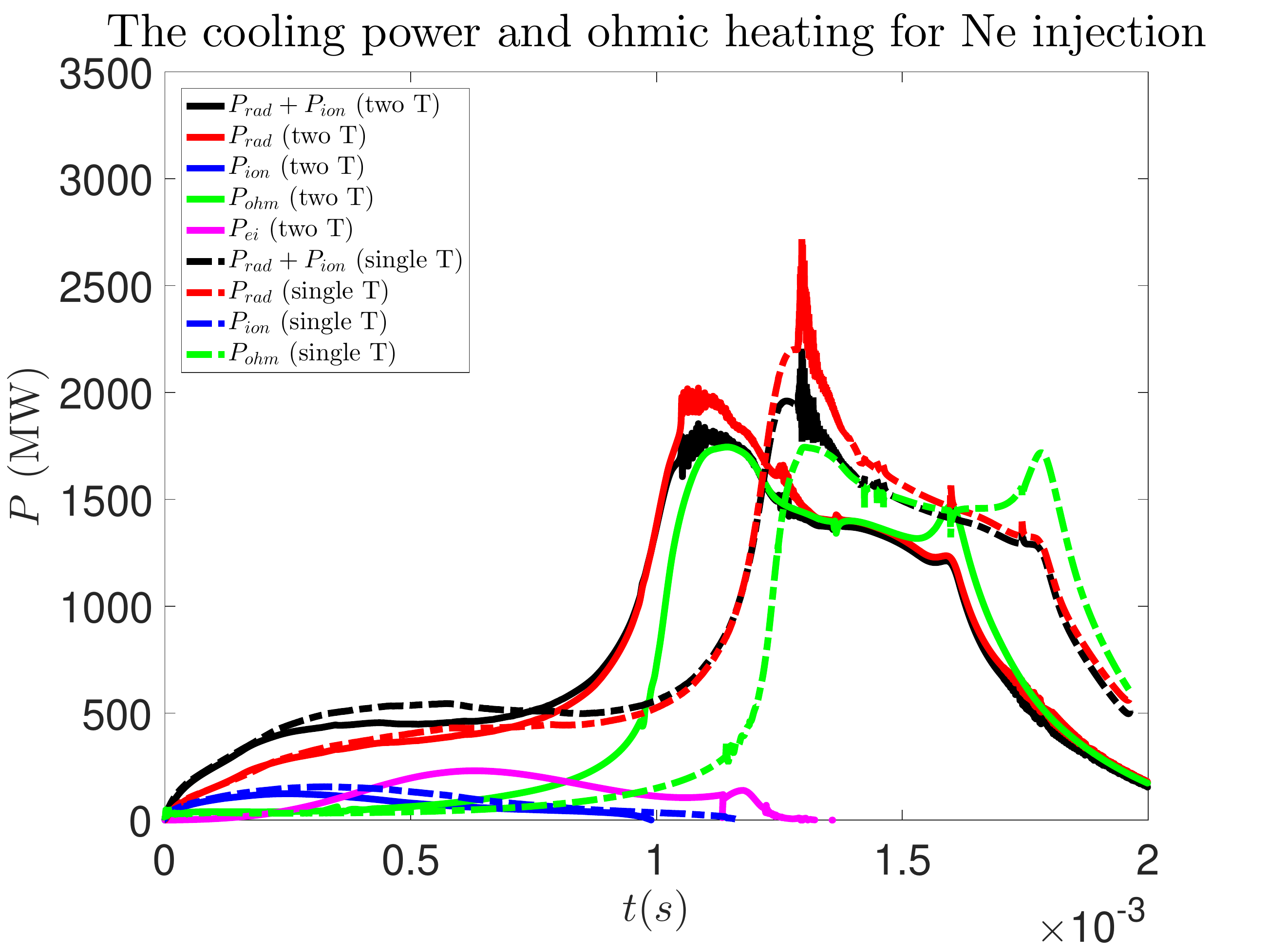}
}
\etbl
\caption{The comparison of JOREK non-equilibrium treatment under the single temperature (dash-dot lines) and the two temperature (solid-lines) case. The red lines are the radiation power, the blue ones the ionization power, the black ones represent the above two combined. The Ohmic power is shown in green, and the inter-specie energy transfer is shown in magenta.}
\label{fig:13}
\end{figure*}

The comparison for the neon injection case is similarly shown in Fig.\,\ref{fig:13}. The solid lines still represent the two temperature model and the dash-dot lines represent the single temperature model. The cooling power shows a plateau before rising into the first peak, which corresponds to the time of plasma cooling down to $100eV$ range. After the ``turning point'' the Ohmic power balances the cooling power, until a second peak marks the termination of the plasma. Once again, the two models agree which each other on the general trend of the time evolution, as well as the peak value of the cooling power and the Ohmic heating power. However, the neon single temperature model exhibits a more pronounced delay than the argon case in all the key characteristic events compared with the two temperature model. This is again due to the fact that the electrons in the two temperature model are cooled down faster in the early phase of the injection, thus reaching the ``turning point'' in the radiation curve faster than their single temperature counterparts.

The ion and electron thermal energy evolution for the above argon and neon injection cases is shown in Fig.\,\ref{fig:14}. The ion thermal energies are shown in blue lines while the electron ones are shown in red lines. The argon injection case is shown with solid lines and the neon one is shown in dash-dot lines. Since the radiative cooling is acting mostly on the electron, one would naively expect to see the electron thermal energy decreasing faster than the ion one. Somewhat anti-intuitively, approximately after $t=0.6ms$ in both cases, the electron thermal energy is higher than the ion one in both cases, this is simply due to the larger particle number of electrons as the electron temperature in general is lower than the ion temperature. Another noteworthy feature is that by the time we have the balance between the cooling power and the Ohmic power in Fig.\,\ref{fig:12} and Fig.\,\ref{fig:13}, the thermal energies from both species are already mostly depleted, and only the Ohmic conversion of the magnetic energy is balancing the radiative cooling.

Overall, the two temperature model shows a faster cooling electron population compared with the single temperature case, although in both cases the peak value of the cooling power and the Ohmic heating power are comparable. Depending on species, the single temperature model could see some time delay in the occurrence of the key characteristic events compared with the two temperature model. This might have some impact on the disruption mitigation behaviors especially considering the interplay between the MHD modes and the fragment locations.

\section{Benchmark with NIMROD \& M3D-C1}
\label{s:Benchmark}

In this section, we provide the benchmark against the published M3D-C1 and NIMROD axisymmetric simulations from Ref. \cite{Lyons2019PPCF}. We look into argon and neon injections with both constant resistivity and Spitzer-like resistivity as explained in Section \ref{s:Eq}. We only use the single temperature model of JOREK in this section.

\subsection{Argon and neon cases with constant resistivity}
\label{ss:ArConstEta}

We begin with argon and neon injections using constant resistivity. We directly extracted the combined radiation and ionization power as a function of time from Ref. \cite{Lyons2019PPCF}. We have chosen to only show the NIMROD data here since the M3D-C1 and the NIMROD data follow each other rather well. We also extracted the total electron number increase due to the injected impurities, as an indication of the ionization state evolution.

\begin{figure*}
\centering
\noindent
\btbl{c}
\parbox{6.5in}{
    \includegraphics[scale=0.45]{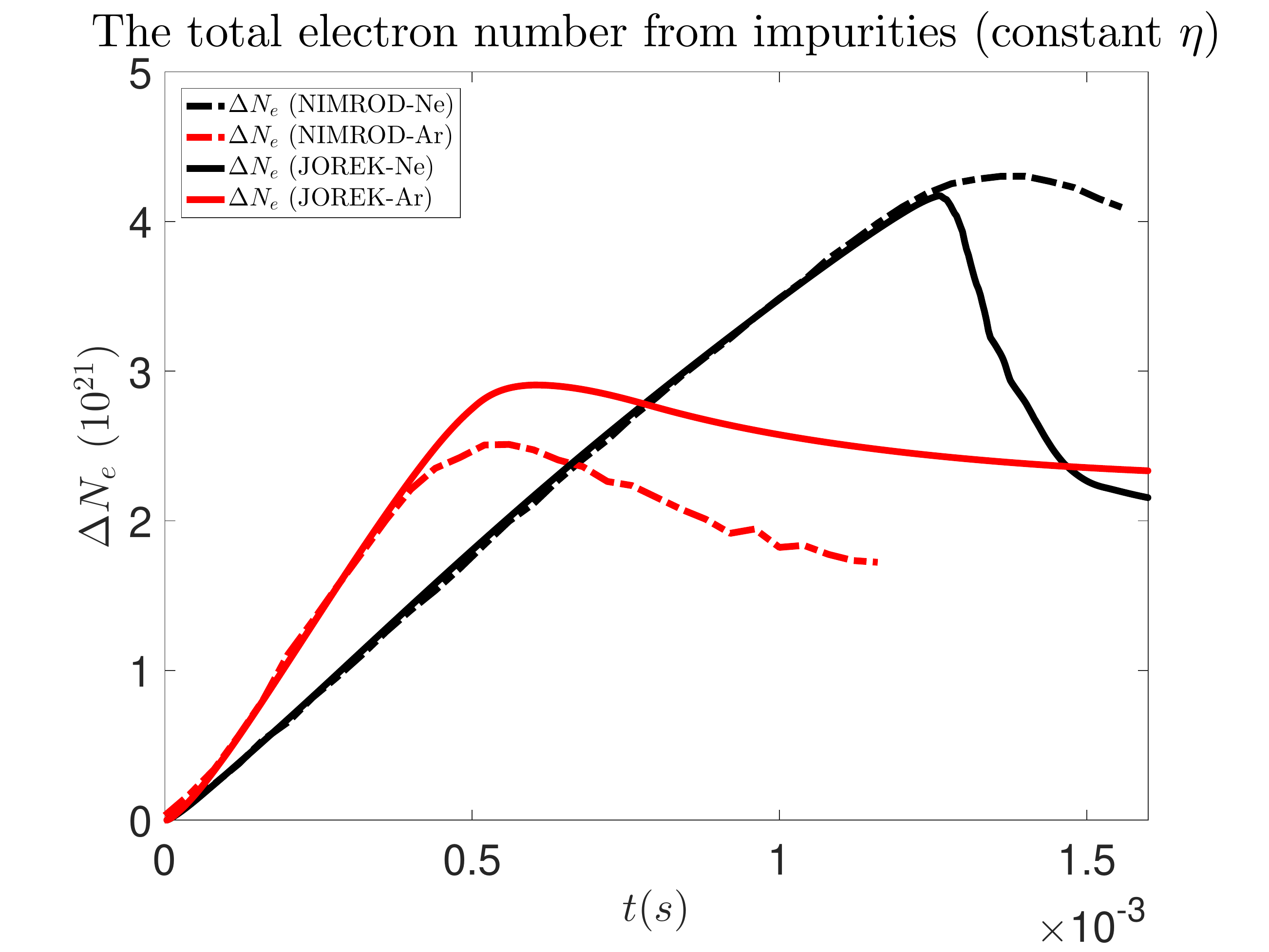}
}
\etbl
\caption{The comparison between JOREK non-equilibrium treatment (solid lines) and the NIMROD data (dash-dot lines) with constant resistivity. The red lines are the argon injection, the black ones are the neon injection.}
\label{fig:04}
\end{figure*}

We compare the total electron number increase between the JOREK and NIMROD results in Fig.\,\ref{fig:04}. The solid lines are the JOREK results while the dash-dot lines are the NIMROD results. The red lines indicates the argon injection results, while the black lines indicate the neon injection results. Very good agreement is obtained in the early phase. However, discrepancies begin to occur near the ``turning points'' of the curves, which correspond to the cooling of the plasma from $100eV$ to $10eV$. We will see later that for argon this is mostly due to a time difference of the plasmas reaching such a temperature region. For the neon case, we will show in \ref{ap:Atomic} that it may be caused by the difference in the recombination probability due to the different atomic models used.

\begin{figure*}
\centering
\noindent
\btbl{c}
\parbox{6.5in}{
    \includegraphics[scale=0.45]{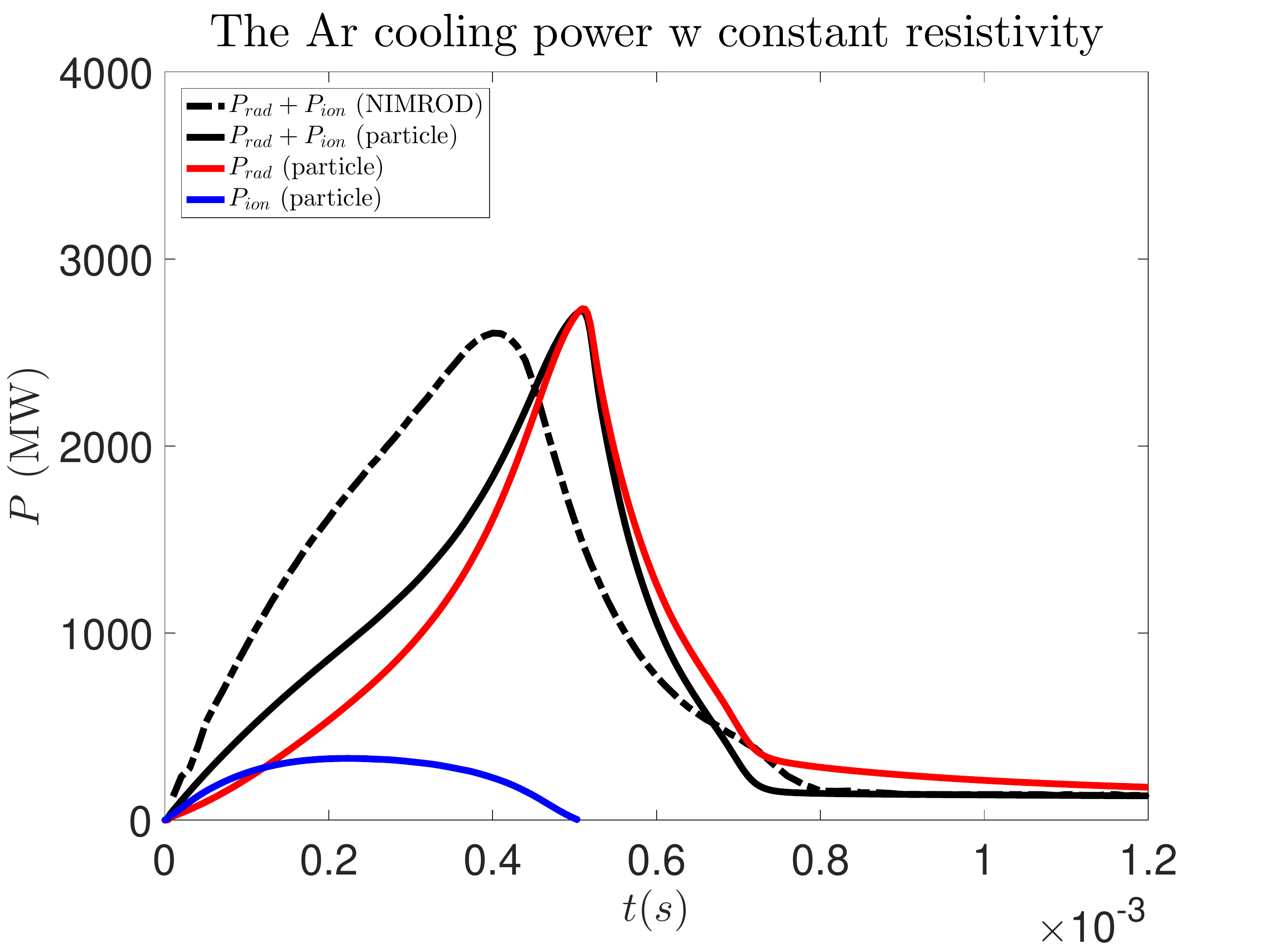}
}
\etbl
\caption{The argon injection comparison between JOREK non-equilibrium treatment (solid lines) and the NIMROD data (dash-dot lines) with constant resistivity. The red solid line is the radiation power, the blue one the ionization power, the black one represents the above two combined. The Ohmic power is not shown due to being constant as the current density change is small.}
\label{fig:05}
\end{figure*}

The cooling power comparison between the JOREK argon injection and the corresponding NIMROD data for the constant resistivity case is shown in Fig.\ref{fig:05}. Comparing the black solid line and the dash-dot line, we can see that the combined radiation and ionization power as a function of time share the same general shape, as well as similar peak level. The JOREK result seems to have its peak delayed by $0.15ms$ compared with the NIMROD result, this is due to the difference in the atomic data used. Indeed, we demonstrate in \ref{ap:Atomic} that much better agreement is achieved if we switch to older versions of the ADAS data in JOREK. Nonetheless, the JOREK case and the NIMROD case deplete the thermal energy by the same time, as is shown by the black solid and the dash-dot lines reaching the constant Ohmic heating level at the same time, showing the robustness of the plasma dynamics. In the later half of the simulation, the red solid line differs slightly from the black solid line. This means that recombination contributes significantly to the radiated power. Despite the deviation in cooling details caused by the difference in the atomic data, the two codes generally agree with each other in both the time evolution of the cooling power curve and the peak cooling power.

\begin{figure*}
\centering
\noindent
\btbl{c}
\parbox{6.5in}{
    \includegraphics[scale=0.45]{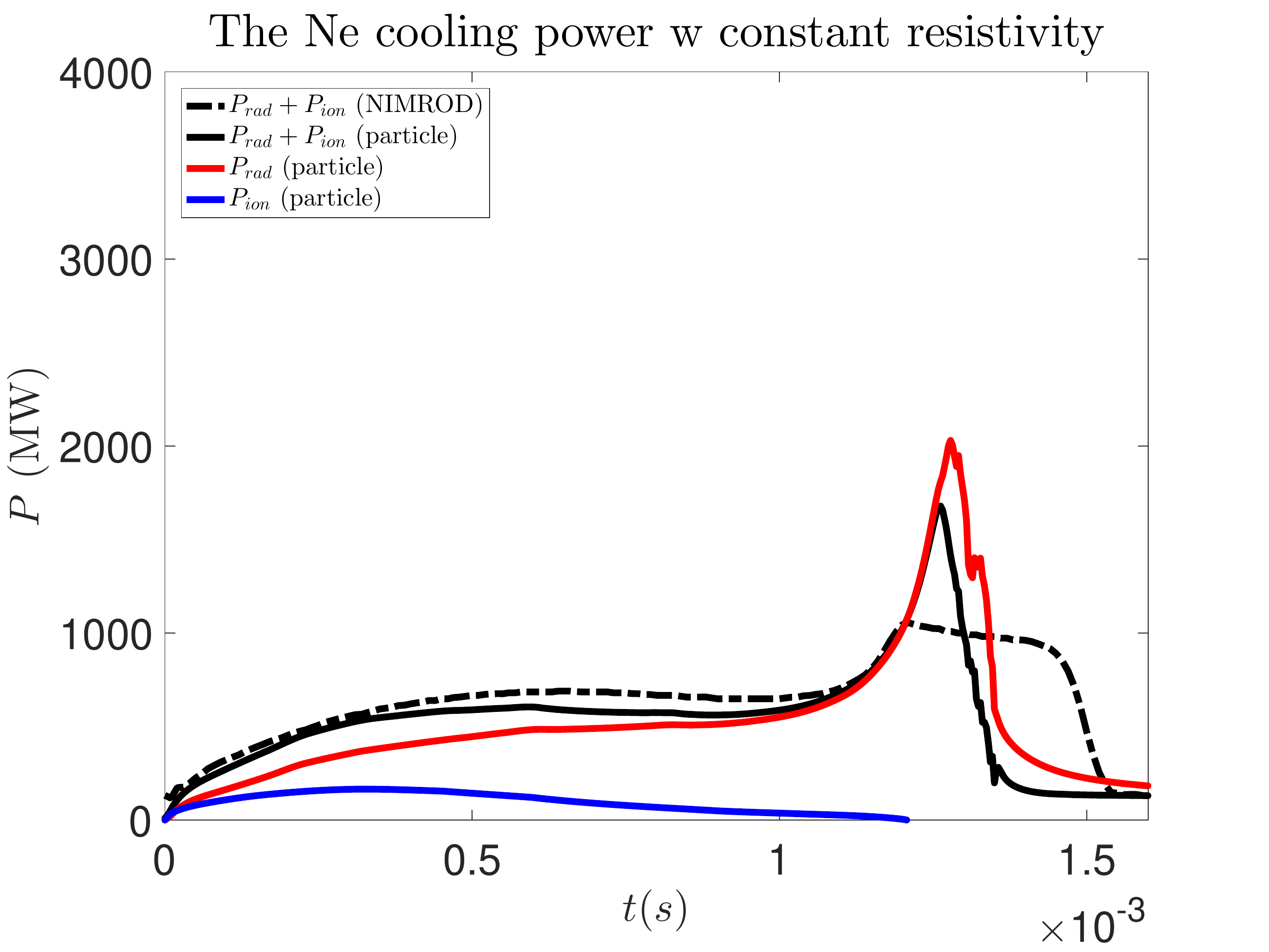}
}
\etbl
\caption{The neon injection comparison between JOREK non-equilibrium treatment (solid lines) and the NIMROD data (dash-dot lines) with constant resistivity. The red solid line is the radiation power, the blue one the ionization power, the black one represents the above two combined. The Ohmic power is not shown due to being constant as the current density change is small.}
\label{fig:06}
\end{figure*}

A similar comparison for neon injections is shown in Fig.\,\ref{fig:06}. Again, the solid lines are the JOREK non-equilibrium treatment results, while the dash-dot lines are the NIMROD results. Very good agreement exists before $t=1.2ms$, but the JOREK result shows a higher radiation peak and subsequent rapid drop of the radiation power, while the NIMROD result plateaus at a lower level for some time. The quick drop of the JOREK radiation power is the result of the rapid recombination shown in Fig.\,\ref{fig:04}, since the radiation power is dependent on the electron density and the charge state distribution as is shown in Eq.\,\rfq{eq:Prad}.

\subsection{Argon and neon case with the Spitzer-like resistivity}
\label{ss:ArSpEta}

\begin{figure*}
\centering
\noindent
\btbl{c}
\parbox{6.5in}{
    \includegraphics[scale=0.45]{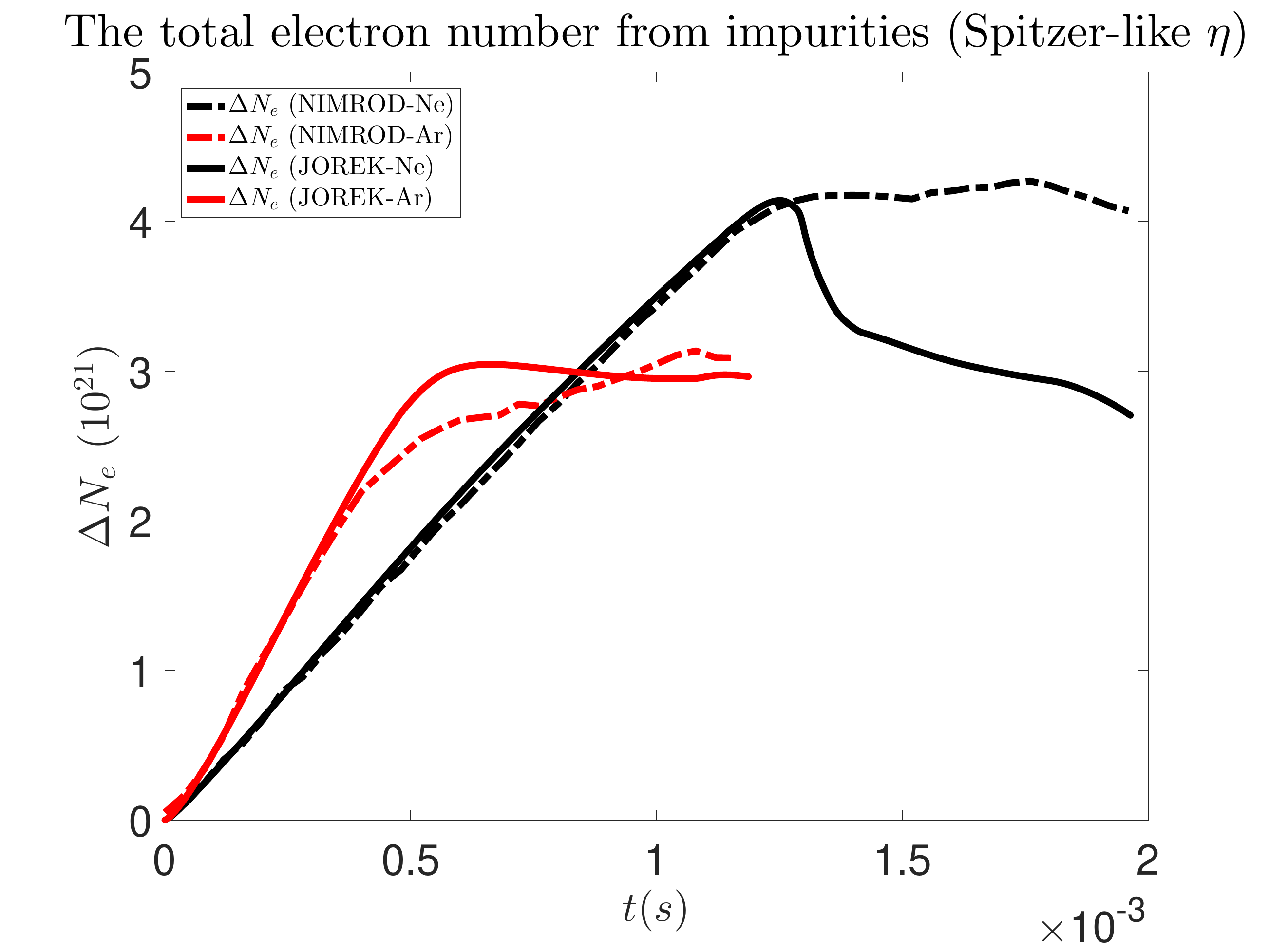}
}
\etbl
\caption{The comparison between JOREK non-equilibrium treatment (solid lines) and the NIMROD data (dash-dot lines) with Spitzer-like resistivity. The red lines are the argon injection, the black ones are the neon injection.}
\label{fig:07}
\end{figure*}

We now compare the argon and neon injection simulated by JOREK and NIMROD with Spitzer-like resistivity. The total electron number from the injected impurities is shown in Fig.\,\ref{fig:07}. Compared with the constant resistivity case, the agreement between the JOREK result and the NIMROD one is better, although the JOREK neon case still shows significant recombination when the plasma core is cooled down below $100eV$ at $t=1.2ms$, corresponding to the turning point in the black solid line. Again, this difference in behaviour is most likely caused by the difference in the atomic data used.

\begin{figure*}
\centering
\noindent
\btbl{c}
\parbox{6.5in}{
    \includegraphics[scale=0.45]{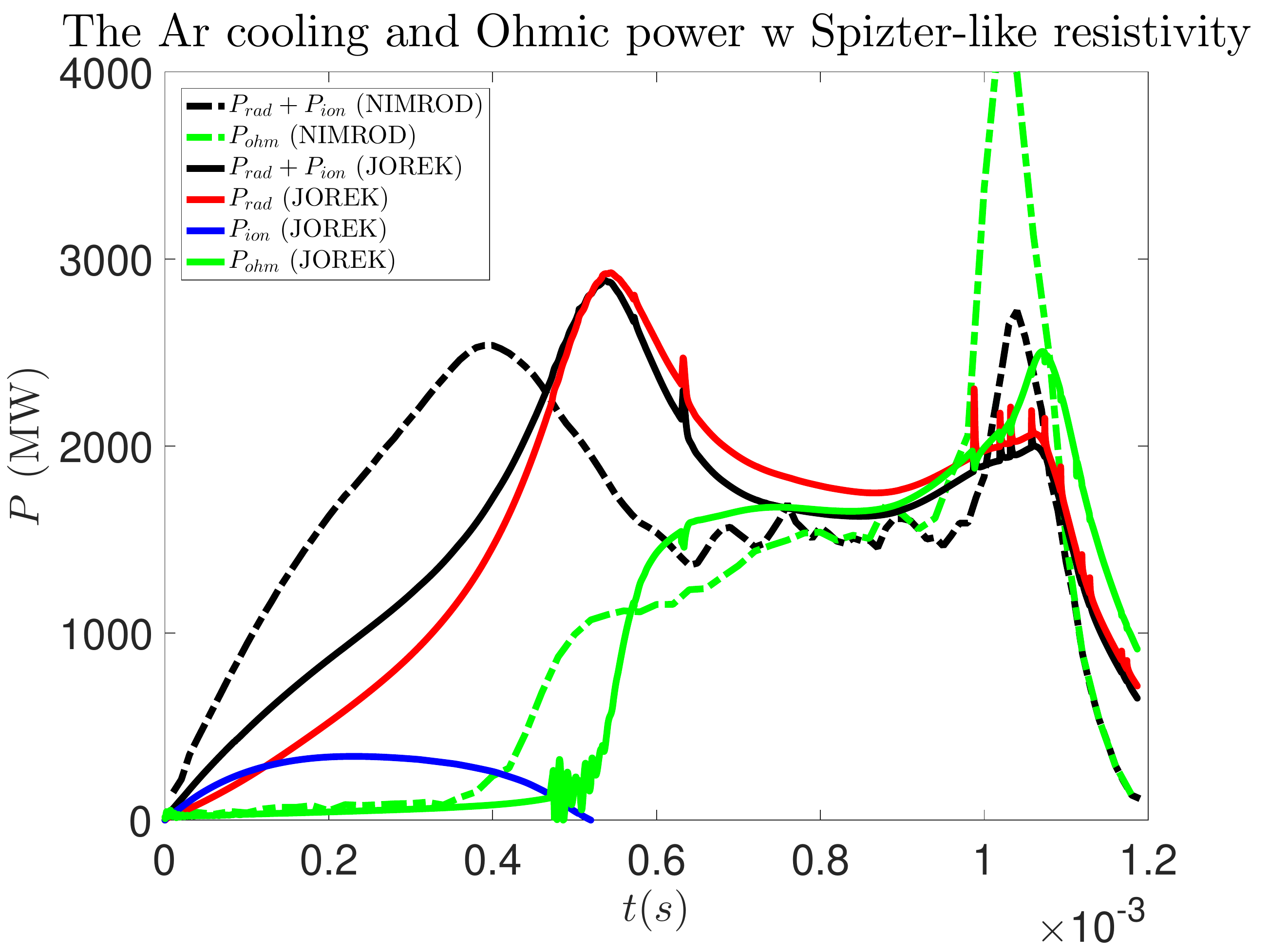}
}
\etbl
\caption{The comparison between JOREK non-equilibrium treatment (solid lines) and the NIMROD data (dash-dot lines) with Spitzer-like resistivity. The red solid line is the radiation power, the blue one the ionization power, the black one represent the above two combined. The Ohmic power for both cases is shown in green solid and dash-dot lines respectively. The JOREK time traces here is the same with the time trace of the non-equilibrium treatment shown in Fig.\ref{fig:02}.}
\label{fig:08}
\end{figure*}

The comparison of the cooling power between the JOREK particle-based treatment and the NIMROD data for the Spitzer-like resistivity case is shown in Fig.\,\ref{fig:08}. Similar to the constant resistivity case, although the cooling peak occurs a bit later for the JOREK case, the two codes generally agree on the peak cooling power and the time evolution. Further, the Ohmic power evolution shows general agreement, and the two codes exhibit the same ``plateau'' behavior on the same power level when the Ohmic power comes to balance the cooling power as the plasma cools down. The two codes also show the same ``bursting'' feature in the Ohmic power and the cooling power at the same time towards the end of the simulation, although in the NIMROD case the Ohmic heating seems to be stronger. It should be noted that such deviation towards the end of the simulation is also present between the NIMROD and M3D-C1 comparison in Ref. \cite{Lyons2019PPCF}, probably due to the difference in the details of the current filament dynamics.

\begin{figure*}
\centering
\noindent
\btbl{cc}
\parbox{2.8in}{
    \includegraphics[scale=0.28]{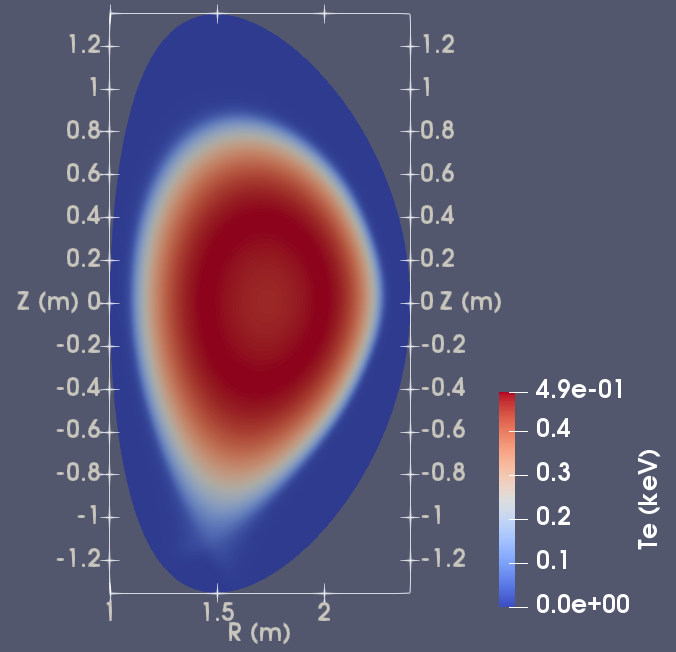}
}
&
\parbox{2.8in}{
	\includegraphics[scale=0.28]{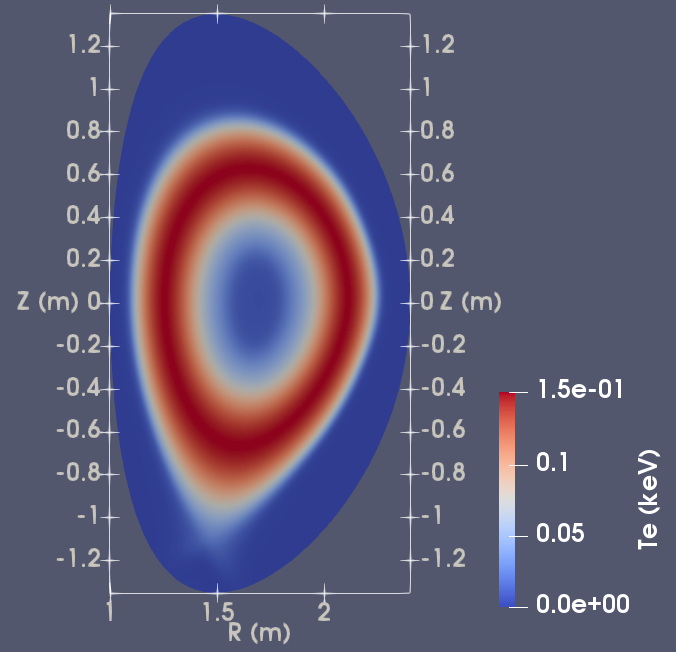}
}
\\
(a)&(b)
\\
\parbox{2.8in}{
  	\includegraphics[scale=0.28]{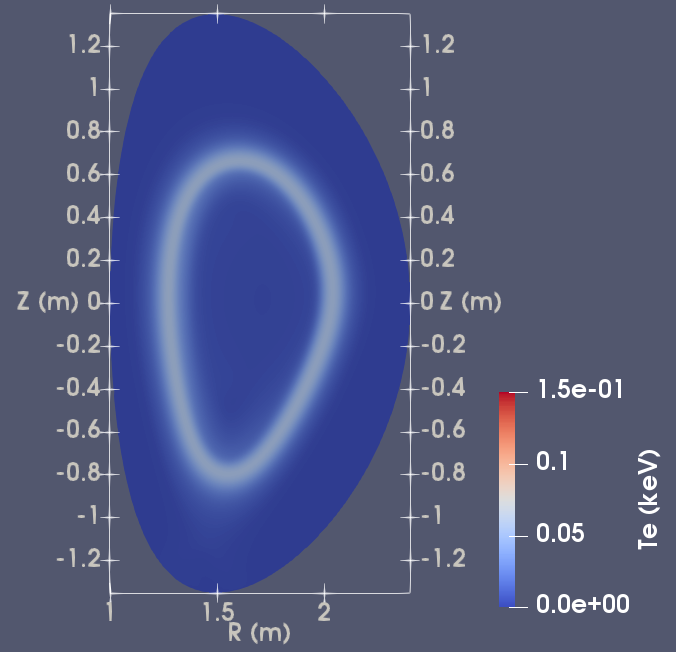}
}
&
\parbox{2.8in}{
	\includegraphics[scale=0.28]{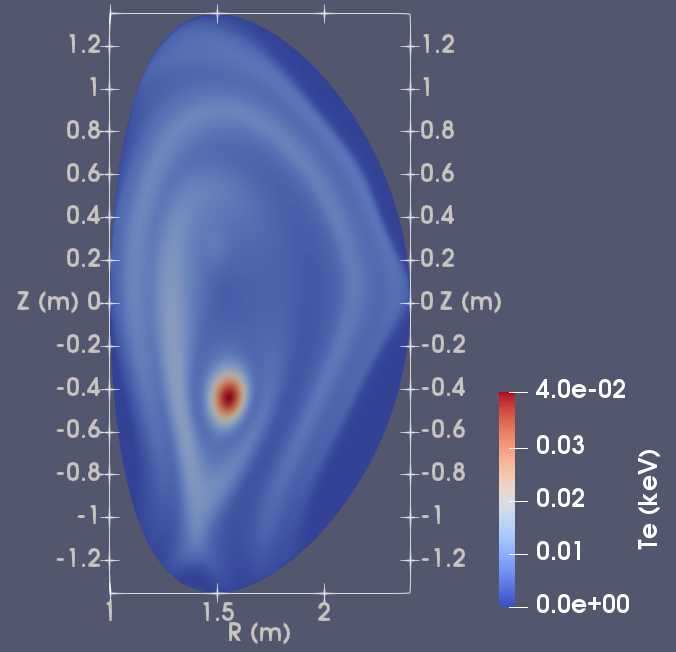}
}
\\
(c)&(d)
\etbl
\caption{The electron temperature evolution during the argon-injection simulated by the JOREK particle-based impurity model with Spitzer-like resistivity at time (a) $t=0.408ms$, (b) $t=0.585ms$, (c) $t=0.890ms$, $t=1.106ms$. Note the different color scale used in some figures.}
\label{fig:09}
\end{figure*}

The agreement can also be seen from the contour evolution of both the temperature and the current density. The electron temperature evolution for the JOREK case with argon injection and Spitzer-like resistivity is shown in Fig.\,\ref{fig:09}.

Compared with Fig.\,7 of Ref. \cite{Lyons2019PPCF}, it can be seen that the JOREK electron temperature evolution follows the same characteristic trend compared with the M3D-C1 and NIMROD results, although the exact timing may slightly differ. From Fig.\,\ref{fig:09}(a) to (b), we see the inside-out core temperature collapse due to the on-axis injection of argon, and a hollow electron temperature profile. This trend of outward cooling continues until there is only a thin electron temperature ``halo'' remaining as shown in Fig.\,\ref{fig:09}(c). After this ``halo'' finally collapsed, we begin to see the re-heating of a very localized portion of the plasma, which is caused by a re-concentration of the toroidal current as will be shown below.

\begin{figure*}
\centering
\noindent
\btbl{cc}
\parbox{2.8in}{
    \includegraphics[scale=0.28]{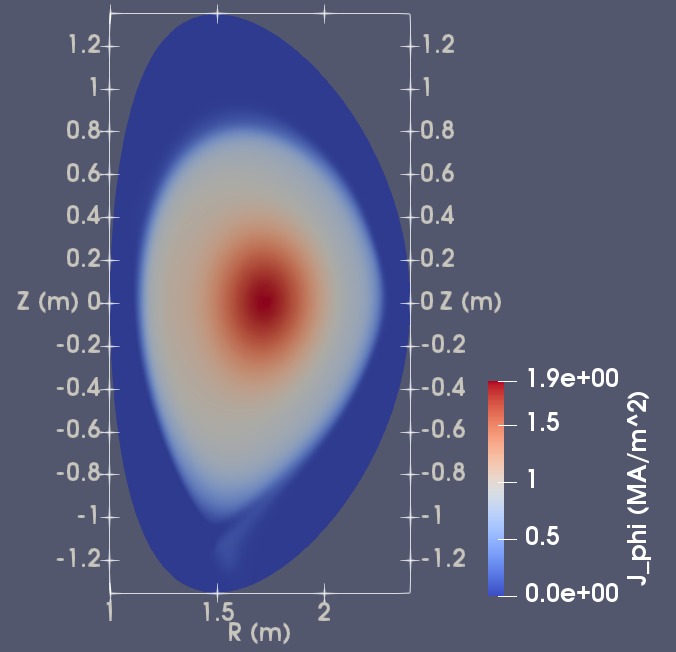}
}
&
\parbox{2.8in}{
	\includegraphics[scale=0.28]{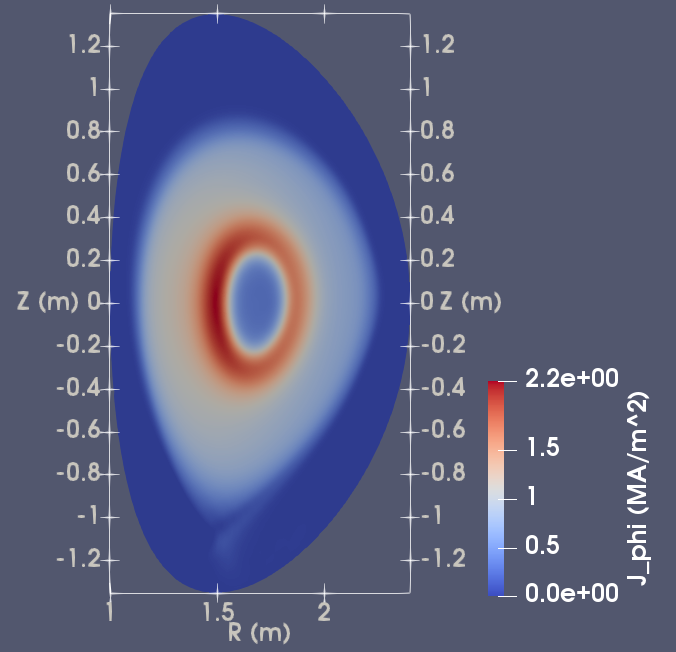}
}
\\
(a)&(b)
\\
\parbox{2.8in}{
  	\includegraphics[scale=0.28]{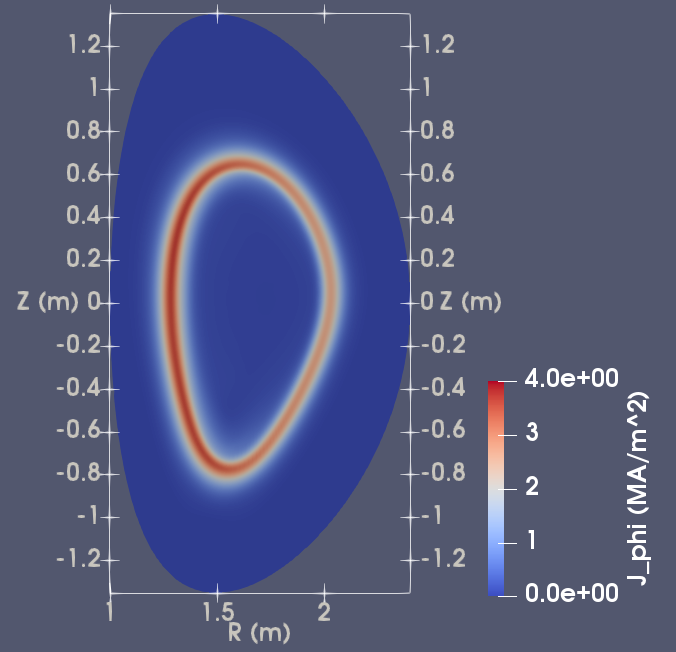}
}
&
\parbox{2.8in}{
	\includegraphics[scale=0.28]{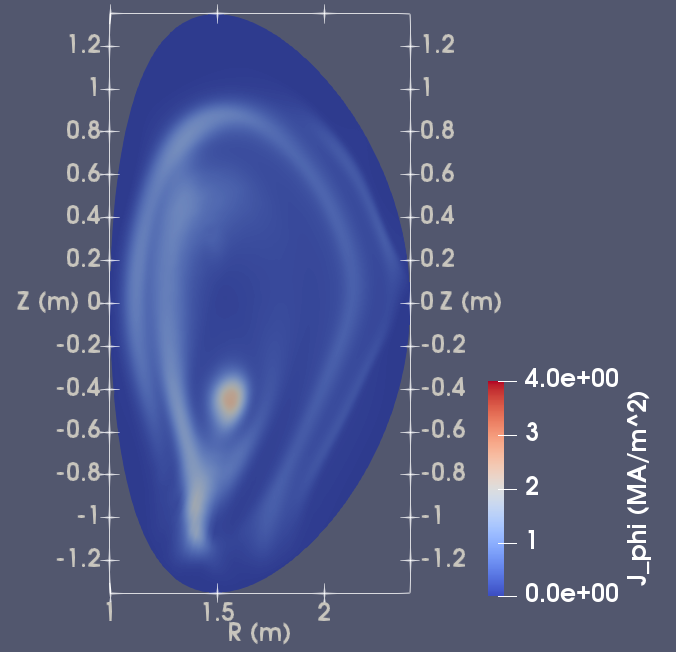}
}
\\
(c)&(d)
\etbl
\caption{The toroidal current density evolution during the argon-injection simulated by the JOREK particle-based impurity model with Spitzer-like resistivity at time (a) $t=0.408ms$, (b) $t=0.585ms$, (c) $t=0.890ms$, $t=1.106ms$. Note the different color scale used in some figures.}
\label{fig:10}
\end{figure*}

The accompanying toroidal current density profile at the same time slices is shown in Fig.\,\ref{fig:10}. Once again, the characteristic trend of the current density profile evolution agrees with the other two codes presented in Ref. \cite{Lyons2019PPCF}, as can be seen by comparing directly with Fig.\,8 of Ref. \cite{Lyons2019PPCF}. From Fig.\,\ref{fig:10}(a) to (b), an inside-out current halo develops as the result of the hollow temperature profile. This current halo in the JOREK case is not as steep as in the other two codes, probably due to the different atomic data used causing the electron temperature profile to differ subtly, thus impacting the details of axisymmetric current redistribution. Nonetheless, as the plasma cools down further, we see the thin current halo forming as shown in Fig.\,\ref{fig:10}(c), exhibiting a similar behaviour as is shown in Ref. \cite{Lyons2019PPCF}. This thin current layer corresponds to the thin electron temperature shell shown in Fig.\,\ref{fig:09}(c), as it is the Ohmic heating that maintain the electron temperature. Last, towards the end of the simulation, we see the re-concentration of the toroidal current near the new magnetic axis as is shown in Fig.\,\ref{fig:10}(d), responsible for the heat spot in Fig.\,\ref{fig:09}(d). This thin current filament can exist only due to the axisymmetric nature of our simulation, otherwise it would terminates by a kink motion due to such a concentration of current density.

\begin{figure*}
\centering
\noindent
\btbl{c}
\parbox{6.5in}{
    \includegraphics[scale=0.45]{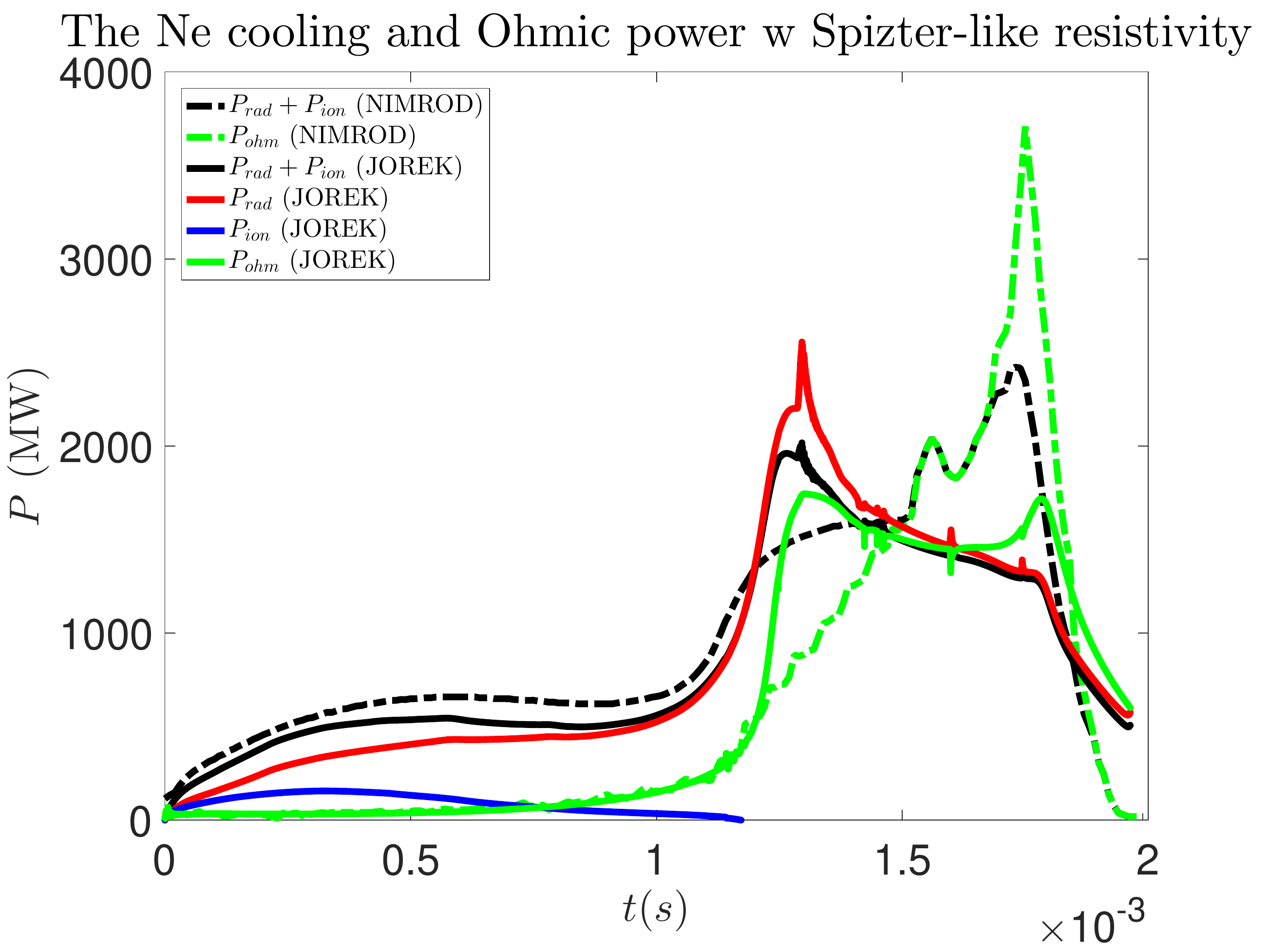}
}
\etbl
\caption{The comparison between JOREK non-equilibrium treatment (solid lines) and the NIMROD data (dash-dot lines) with Spitzer-like resistivity. The red solid line is the radiation power, the blue one the ionization power, the black one represents the above two combined. The Ohmic power for both cases is shown in green solid and dash-dot lines respectively.}
\label{fig:11}
\end{figure*}

Like-wise, the cooling power comparison of the neon injection is shown in Fig.\,\ref{fig:11}. The solid lines indicate the JOREK result while the dash-dot lines indicate the NIMROD results. Good agreement is once again found in the early to middle phase of the simulation, but the JOREK result still shows somewhat higher radiation peak when the plasma is cooled down below $100eV$ at $t=1.3ms$. Nevertheless, the plasma termination occurs approximately at the same time for the two codes, as can be seen by comparing the peak time of the two green curves near $t=1.7ms$.

Overall, despite the difference in the atomic models used, the JOREK particle-based impurity model shows reasonable agreement with M3D-C1 and NIMROD simulations both in terms of the time evolution of integrated quantities and in terms of the characteristic 2D profile evolution of the current density and the temperature. Apart from demonstrating the capability of the JOREK non-equilibrium impurity treatment, such agreement also highlights the robustness of the radiative cooling dynamics.

\section{Conclusion and discussion}
\label{s:Conclusion}

In this paper, we have introduced the new particle-based collisional-radiative non-equilibrium impurity treatment in JOREK. This new treatment is based on the coupled evolution of marker-particles and the fluid fields. The marker-particles are generated according to the impurity density source term, and are pushed along the fluid velocity field lines while ionizing and recombining independently according to local plasma parameters. We then project the particle contribution of the ionization power, the radiative power, the effective and mean charge and the impurity density back to the fluid evolution.

To illustrate the performance of this new non-equilibrium model, we presented a series of comparison and benchmark cases both against the CE result and against M3D-C1 and NIMROD non-equilibrium impurity injection simulations. As for the comparison against the JOREK CE treatment using Spitzer-like resistivity, we find the particle-based model successfully catches the early cooling dynamics which were missed by the previous CE treatment, although the peak radiation value is comparable between the non-equilibrium and the CE treatment.

Under the single temperature model, the CE treatment missed the quick rise of the Ohmic heating power in the middle of the simulation, as it underestimates the early phase cooling and did not bring the electron temperature down fast enough. Only towards the end of the simulation, we see the balance between the Ohmic heating and the radiative cooling. Under the two temperature treatment, the CE model behaves better as the peak radiative power, the peak Ohmic power and the power balance are comparable with the non-equilibrium model, although the CE result still show some delay in time.

As a demonstration, we also show the evolution of the non-equilibrium mean charge number profile and the comparison with its CE value using the non-equilibrium, two temperature argon injection case. The mean charge number generally show some delay in the time evolution due to the finite time requirement to reach the CE state, highlighting the importance of the non-equilibrium description when describing the impurity dynamics in a rapidly changing plasma background.

Comparing the non-equilibrium impurity behavior between the single and the two temperature treatment, it is found that the electrons in the two temperature model generally are cooled down faster compared with their single temperature counterparts. This results in earlier ``turning point'' in the radiation power curve, although the peak cooling power is comparable between the single and the two temperature models. This time difference is especially pronounced in the neon injection case. Such comparison has shown that the two temperature model is capable of capturing the temperature deviation between species at the time of the radiative collapse. Such deviation could be important for the MHD and the ablation dynamics during a DMS scenario.

Moreover, compared with the coronal non-equilibrium simulation of M3D-C1 and NIMROD, the JOREK non-equilibrium result shows good agreement with the other two codes in terms of the peak cooling power, the Ohmic power, the power balance and the time evolution of those quantities. For the argon case, the time by which the first radiation peak is reached differs slightly between the JOREK result and that of the other two codes, due to the difference in the atomic data used. For the neon case, the JOREK results show stronger radiative power and recombination when the plasma is cooled down below $100eV$. Despite these deviations, the later plasma evolution and the plasma termination process show the same characteristic behaviours among the three codes, highlighting the robustness of the plasma dynamics.

Since we are only pushing the marker particles along the velocity field lines thus only taking into consideration the convective transport of the impurities, our current method may suffer some inconsistency in describing the impurity density evolution should there be a strong impurity diffusion. However, since our current scenarios of interest are mostly dominated by convective transport, this is not immediately foreseen to be problematic. We are working on an improved particle pusher to take into account the diffusive transport of the marker particles so that we can later on expand our study to cases with strong impurity diffusion.

Overall, we have shown that the JOREK non-equilibrium impurity model is now ready for production. It is capable of accurately capturing the early injection phase cooling and is compatible with both the single and the two temperature model, paving the way for future simulation of important DMS scenarios.

\appendix
\section{The impact of atomic data onto the simulation result}
\label{ap:Atomic}

\begin{figure*}
\centering
\noindent
\btbl{c}
\parbox{6.5in}{
    \includegraphics[scale=0.45]{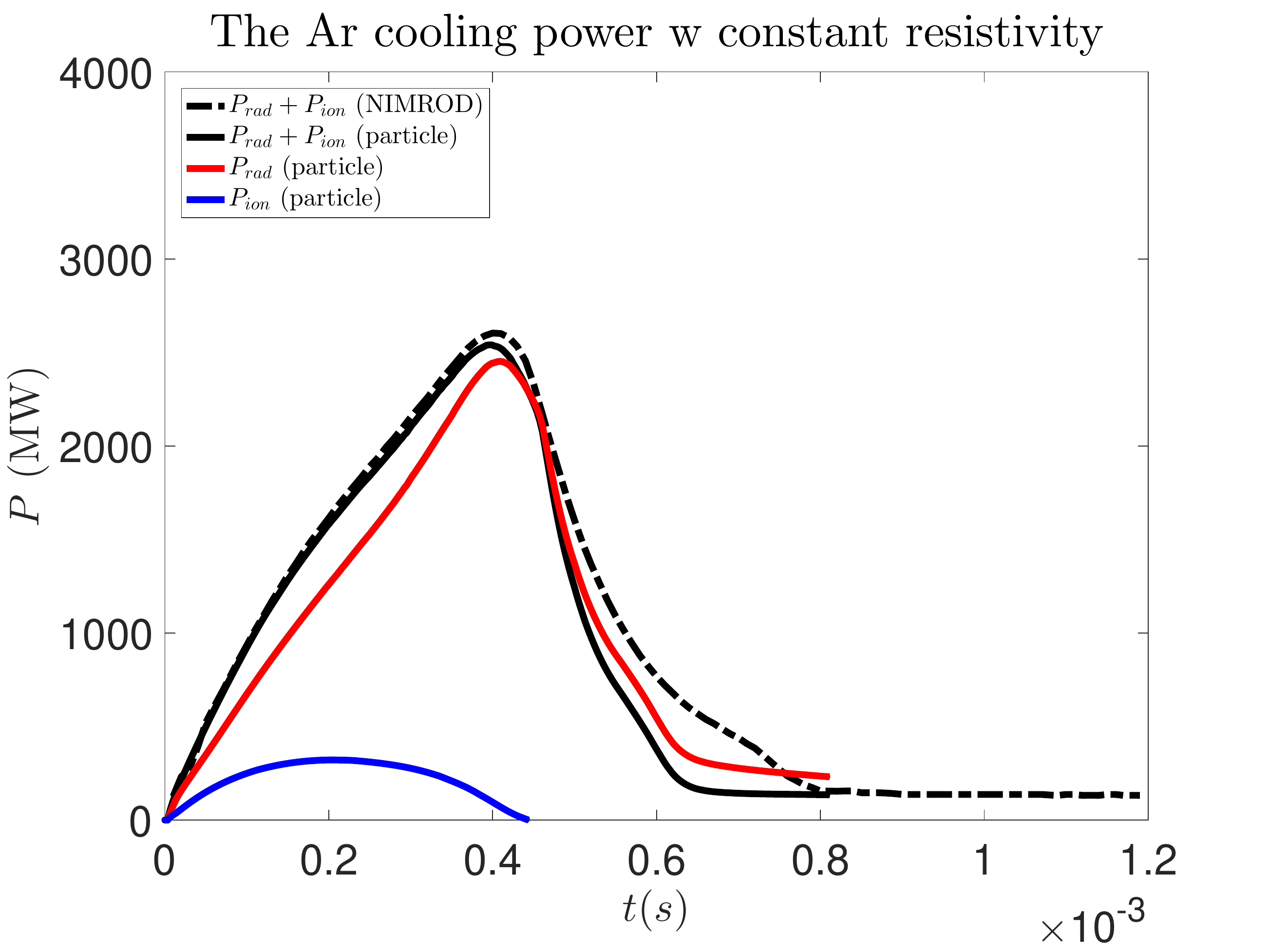}
}
\etbl
\caption{The argon injection comparison between JOREK non-equilibrium treatment (solid lines) and the NIMROD data (dash-dot lines) with constant resistivity. The 1942 argon line radiation data is used instead of the default 1989 version. The red solid line is the radiation power, the blue one the ionization power, the black one represents the above two combined. The Ohmic power is not shown due to being constant.}
\label{fig:ap01}
\end{figure*}

As mentioned in the main article, the choice of the atomic data may have some impact on the details of the cooling power time evolution, although the plasma dynamics ultimately is shown to have some robustness. Here, we will demonstrate that by using an older version of the ADAS data we can recover better agreement with the NIMROD and the M3D-C1 result for the argon case, thus supporting our claim that the deviation shown in Section \ref{s:Benchmark} are indeed due to the different atomic data used. For neon, we don't find a good set of atomic data by which we could exactly reproduce the NIMROD and M3D-C1 results. Instead, we will give a comparison of the line radiation, the ionization and the recombination coefficient between the ADAS and the KPRAD fitting of the ADPAK data \cite{Lyons2019PPCF,LyonsKim}, and demonstrate that the discrepancy between JOREK and the other two codes near the plasma termination can indeed be explained by the difference in the atomic data used.

We hereby rerun the argon injection benchmark case with constant resistivity first shown in Section \ref{ss:ArConstEta}, but this time we use the ADAS argon line radiation data of 1942 instead that of 1989. This is the only change we do and all other data are the same as what is used in Section \ref{ss:ArConstEta}. The comparison between the JOREK non-equilibrium impurity result and the NIMROD result is shown in Fig.\,\ref{fig:ap01}.

Like in previous figures, the solid lines are the JOREK result and the dash-dot lines are the NIMROD result. The black lines are the combined cooling power, while the red line is the radiation power and the blue line is the ionization power. Comparing Fig.\,\ref{fig:ap01} to Fig.\,\ref{fig:05}, the agreement between the JOREK result and the NIMROD result is much better when we use the old 1942 version ADAS data. The slight deviation after $t=0.4ms$ shown in Fig.\,\ref{fig:ap01} is likely due to the fact that we are still using a newer version of the argon recombination and continuum radiation data, since the 1942 version of such data is not available. Nevertheless, Fig.\,\ref{fig:ap01} confirms our previous claim that the previous deviation in Section \ref{s:Benchmark} are indeed mainly due to the different atomic data used.

\begin{figure*}
\centering
\noindent
\btbl{c}
\parbox{4.0in}{
    \includegraphics[scale=0.65]{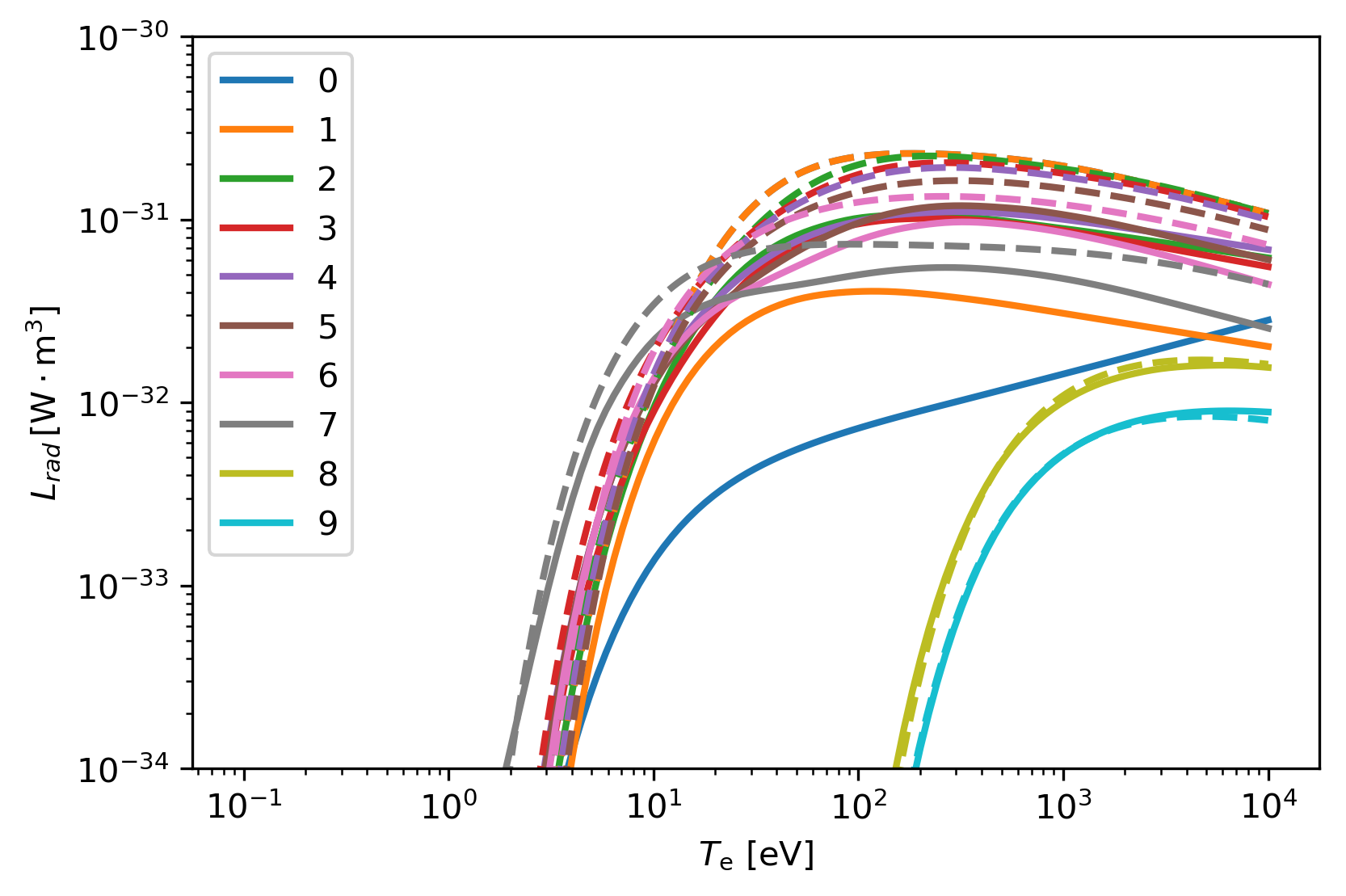}
}
\\
(a)
\\
\parbox{4.0in}{
	\includegraphics[scale=0.65]{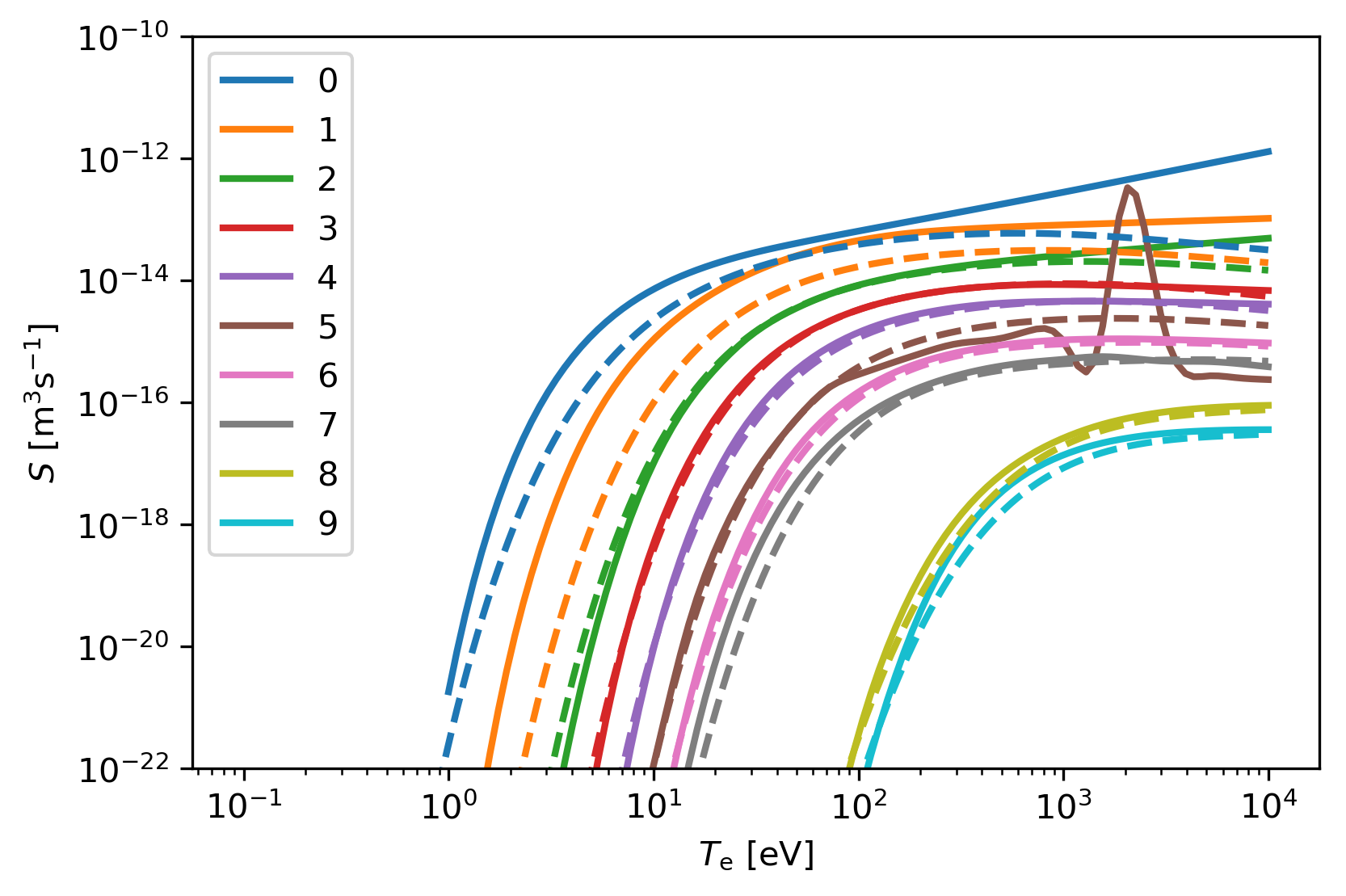}
}
\\
(b)
\\
\parbox{4.0in}{
	\includegraphics[scale=0.65]{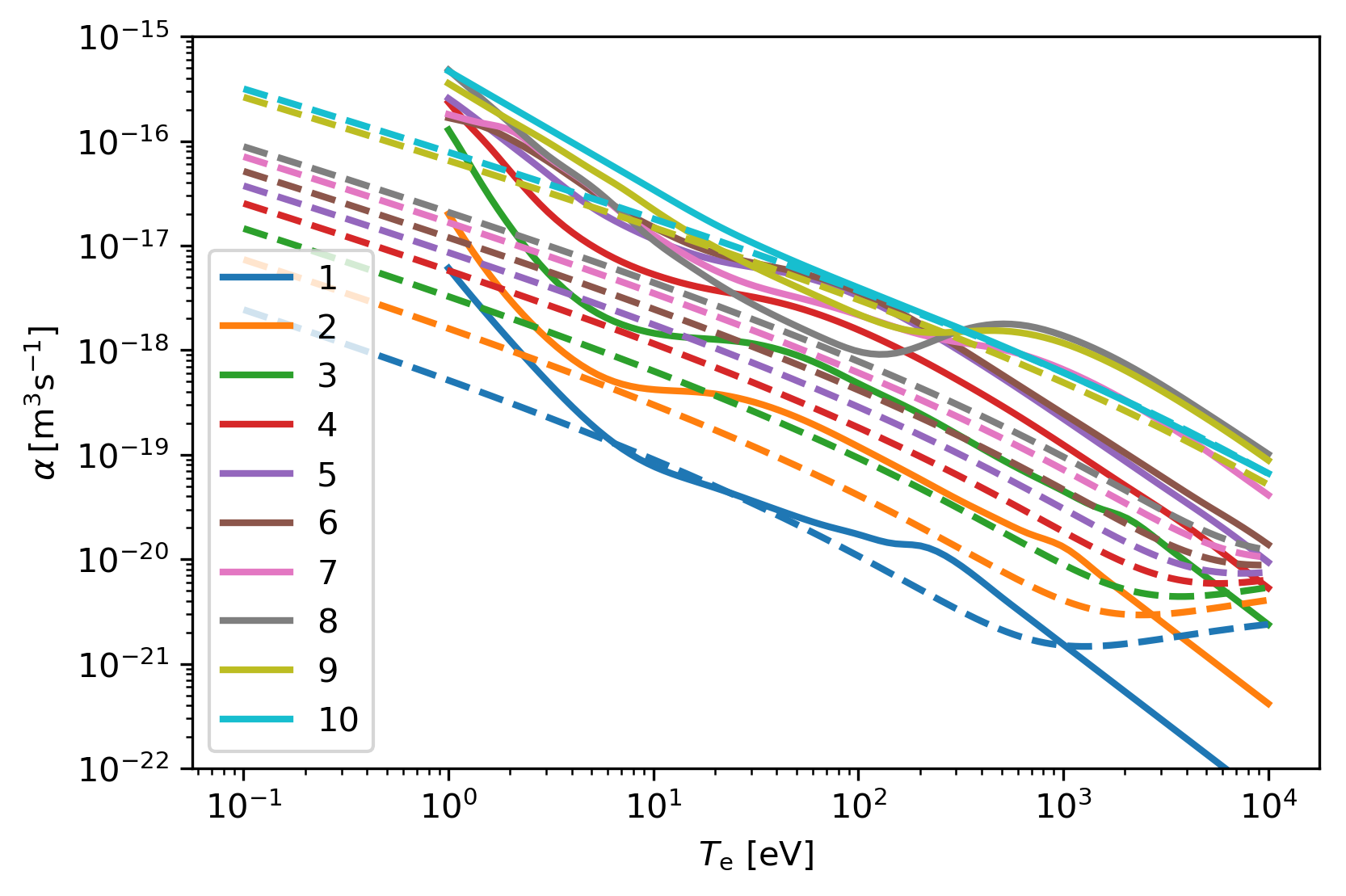}
}
\\
(c)
\etbl
\caption{The (a) line radiation power function $L_{rad,L}$, (b) ionization rate $S$ and (c) recombination rate $\ga$ for each neon charge state in the ADAS (solid line) and the fitted ADPAK (dashed line) data.}
\label{fig:ap02}
\end{figure*}

For the neon case, the comparison of the line radiation, the ionization and the recombination coefficient between the ADAS data and the fitted ADPAK data at $n_e=10^{21}/m^3$ is shown in Fig.\,\ref{fig:ap02}. For the line radiation and the ionization coefficient shown in Fig.\,\ref{fig:ap02}(a) and (b), it could be seen that the agreement at strongly ionized states are quite good between the two atomic data sets, but for the weakly ionized states at high temperature there could be significant difference. More importantly, the recombination for the charge states $4+$ to $8+$ could have order of magnitude difference depending on the electron temperature, with the ADAS data exhibiting stronger recombination in general. This is consistent with our observation of faster neon recombination when the plasma is cooled down as is shown in the ``turning point'' of the electron number in Fig.\,\ref{fig:04}. The discrepancy in the radiated power between the JOREK result and the NIMROD result around $t=1.3ms$ in Fig.\,\ref{fig:06} could also be explained by the difference of the charge state distribution as the consequence of the the aforementioned difference in the recombination coefficient.

As conclusion, we acknowledge that difference in the atomic data could cause observable impact onto the details of radiation collapse dynamics, providing incentives for further detailed studies in the future to determine what is the best set of atomic data to be used. We would also like to emphasize again that despite the difference in the atomic data, our JOREK result still shows reasonable agreement in terms of the peak cooling power, Ohmic power, 2D profile evolution and characteristic events, highlighting the robustness of the plasma dynamics in the massive material injection scenarios we have been studying.

\vskip1em
\centerline{\bf Acknowledgments}
\vskip1em

  The authors thank B.C. Lyons and C.C. Kim for providing the KPRAD fitting of the ADPAK atomic data and for their constructive suggestions. We also thank D. van Vugt, B. Nkonga, S.J. Lee for fruitful discussion. ITER is the Nuclear Facility INB no. 174. The views and opinions expressed herein do not necessarily reflect those of the ITER Organization. This publication is provided for scientific purposes only. Its contents should not be considered as commitments from the ITER Organization as a nuclear operator in the frame of the licensing process. Part of this work is supported by the National Natural Science Foundation of China under Grant No. 11905004, and the National MCF Energy R\&D Program of China under Grant No. 2019YFE03010001. This work has been co-funded by the ITER Organization under the implementing agreement IO/IA/19/4300002026. Part of this work has been carried out within the framework of the EUROfusion Consortium and has received funding from the Euratom research and training program 2014-2018 and 2019-2020 under grant agreement No 633053. The views and opinions expressed herein do not necessarily reflect those of the European Commission. This work is carried out on Tianhe-3 prototype operated by NSCC-TJ.

\end{document}